\documentclass[prd,aps,twocolumn,a4paper,floatfix,nofootinbib]{revtex4}

\usepackage{graphicx,psfrag}
\usepackage{mathrsfs}
\usepackage{amsmath,amsfonts,amssymb}
\usepackage{multirow,hyperref}
\usepackage{comment}
\usepackage{enumerate}
\usepackage{mathrsfs}
\usepackage{pifont}
\usepackage[normalem]{ulem}
\usepackage{subfigure}
\usepackage[utf8]{inputenc} 

\newcommand{\be}{\begin{equation}}
\newcommand{\ee}{\end{equation}}
\newcommand{\bea}{\begin{eqnarray}}
\newcommand{\eea}{\end{eqnarray}}
\newcommand{\bel}{\begin{align}}
\newcommand{\eel}{\end{align}}

\newcommand{\bse}{\begin{subequations}}
\newcommand{\ese}{\end{subequations}}

\def\GMc2{{\rm G M_{\odot} c^{-2}}}

\usepackage{color}
\definecolor{cyan}{rgb}{0,0.9,0.9}
\definecolor{orange}{rgb}{0.9,0.5,0}
\definecolor{magenta}{rgb}{1,0,1}
\definecolor{purple}{rgb}{0.8,0.4,0.8}
\definecolor{gray}{rgb}{0.8242,0.8242,0.8242}

\begin{document}

\title{Increasing the Accuracy of Binary Neutron Star
       Simulations with an improved Vacuum Treatment}
\author{Amit Poudel$^1$}
\author{Wolfgang Tichy$^1$}
\author{Bernd Br\"ugmann$^2$}
\author{Tim Dietrich$^3$}

\affiliation{${}^1$ Department of Physics, Florida Atlantic University, Boca Raton, FL 33431 USA}
\affiliation{${}^2$ Theoretical Physics Institute, University of Jena, 07743 Jena, Germany}
\affiliation{${}^3$ Institute of Physics and Astronomy, University of Potsdam,
14476 Potsdam, Germany}

\date{\today}

\begin{abstract}
   Numerical relativity simulations are essential to study the last stages
   of the binary neutron star coalescence. Unfortunately, for stable simulations there is the need
   to add an artificial low-density atmosphere. Here we discuss a new framework in which we can effectively
   set the density surrounding the neutron stars to zero to ensure a more accurate simulation.
   We test our method with a number of single star test cases and for an equal mass binary neutron star simulation.
   While the bulk motion of the system is not influenced, and hence, there is no improvement with respect to the
   emitted gravitational-wave signal, we find that the new approach is superior with respect to mass conservation
   and it allows a much better tracking of outward moving material.
   This will allow a more accurate simulation of the ejected material and supports
   the interpretation of present and future multi-messenger observations with more accurate
   numerical relativity simulations.
\end{abstract}


\maketitle

\section{Introduction}
\label{sec:intro}


Binary neutron star (BNS) systems are a unique laboratory to answer some of
the most interesting questions in modern physics. For example: What is the
equation of state (EOS) of supranuclear dense
matter~\cite{Annala:2017llu,Abbott:2018exr,De:2018uhw,Most:2018hfd,Coughlin:2018fis,
Capano:2019eae,Radice:2018ozg,Essick:2020flb,Dietrich:2020lps}?
What is the expansion rate of the
Universe~\cite{Schutz:1986gp,Abbott:2017xzu,Coughlin:2019vtv,Dhawan:2019phb,
Coughlin:2020ozl}?
How have the heavy elements in the
Cosmos~\cite{Cowperthwaite:2017dyu,Smartt:2017fuw,Kasliwal:2017ngb,
Kasen:2017sxr,Watson:2019xjv} been produced?
And, is General Relativity the correct theory to describe
gravity~\cite{GBM:2017lvd,Ezquiaga:2017ekz,Baker:2017hug,Creminelli:2017sry}?

An investigation of the full BNS coalescence requires a
detailed analysis and understanding of the merger process.
Due to the strong gravitational fields and the high velocities of
the stars just before merger, one has to solve Einstein's Equations
with all nonlinearities using full 3+1D numerical-relativity
simulations~\cite{Alcubierre_book,Baumgarte_Shapiro_book,Rezzolla:2013}.
Thus, numerical relativity has consolidated its role for the
interpretation of compact binary mergers and was used to study the
BNS merger GW170817~\cite{TheLIGOScientific:2017qsa}
and its electromagnetic counterparts~\cite{Monitor:2017mdv}.

To enable stable simulations, state-of-the-art numerical relativity simulations 
of neutron stars use an artificial atmosphere to model vacuum and near-vacuum 
conditions outside the stars, see
e.g.~\cite{Yamamoto:2008js,Thierfelder:2011yi,Rezzolla:2013,Baiotti:2016qnr}.
Starting with initial data for neutron stars in vacuum, 
the standard method fills all the vacuum regions with a very
low-density atmosphere (with often a cold equation of state). This
atmosphere is not physical and artificially added for numerical reasons. One
reason for this approach is that for the matter evolution we use conserved
matter variables, i.e.\ variables whose change inside a given cell volume is
determined by fluxes across the cell surfaces. To compute these fluxes one
has to use interpolation from the cell centers to cell interfaces. In low
density regions this interpolation can return matter densities or energies
that lie outside what is physically reasonable or allowed. An artificial
atmosphere cures these issues.
However, even with an artificial atmosphere, some of the same problems can
still occur. In addition, the atmosphere has to be tuned to avoid most of
these problems, while at the same time keeping it tenuous enough to not
unduly influence the simulation.
One of the most sophisticated atmosphere implementations is explained
in~\cite{Radice:2013xpa}. In this approach a positivity preserving limiter
is used for the density. Yet even in this approach a low density atmosphere
is still needed. However, it has the advantage that the density of the
atmosphere can be made much lower than in more straightforward approaches,
so that the effects of the artificial atmosphere can be reduced.
There is also a new hydrodynamics approach that uses Hamilton-Jacobi
methods~\cite{Westernacher-Schneider:2019uyx}, and thus its evolution
equations take a different form. So far it has been only used
for barotropic fluids, and interestingly for us, in its current formulation
it also requires an artificial atmosphere.

Our goal here is to find a scheme that does not explicitly add such an
atmosphere. We will first describe the ingredients that allow us to perform
simulations that contain true vacuum. After this we discuss tests of our new
scheme, where we evolve neutron stars with and without atmosphere.

Throughout this study, we use dimensionless units where $G$=$c$=$M_\odot$=1.
and adopt the signature $(-,+,+,+)$ for the 4-metric.
Greek indices on tensors run from 0 to 3, Latin indices from 1 to 3,
with the standard summation convention for repeated indices.
The following can be used to convert from dimensionless units to SI units:
time 1000 = $4.93 ms$,
distance 1 = $1.47735 km$,
energy 1 = $1.7872\cdot10^{47}J$
and density 1 = $6.177413\cdot10^{17} g/cm^3$

\section{Numerical Method}
\label{sec:method}

\subsection{The BAM code}
\label{sec:method:BAM}

We perform our dynamical simulations with the BAM code~\cite{Bruegmann:2006at,Thierfelder:2011yi,
Dietrich:2015iva,Bernuzzi:2016pie,Dietrich:2018bvi}, which uses the method-of-lines with
Runge-Kutta time integrators and finite differences approximating spatial
derivatives. A Courant-Friedrich-Lewy
factor of~$0.25$ is employed for all runs (see~\cite{Bruegmann:2006at,Cao:2008wn}).

The numerical domain contains a mesh made of a hierarchy of cell-centered
nested Cartesian boxes and consists of $L$ refinement levels
from $l = 0$ to $L-1$. Each refinement level is made out of one
or more equally spaced Cartesian grids with grid spacing $h_l$. There are
$n$ points per direction on each grid plus a certain number of buffer
points on each side.
The levels are refined in resolutions by a factor of two such that the grid
spacing in level $l$ is $h_l = h_0/2^l$,
where $h_0$ is the grid spacing of the coarsest level.
The coordinate extent of a grid at level $l\geq 0$ entirely contains grids
at any level greater or equal to $l+1$.
The moving boxes technique is used to dynamically move and adapt some
of the mesh refinement levels during the time evolution.
These moving refinement levels are used for the cases like
BNS where the
center of each star moves during the time evolution.
All levels with $l>l_{\rm m}$ are moving refinement levels. This is implemented
in such a way that the moving refinement levels always stay within the
coarsest level.
The number of points in one direction for moving level ($n_{\rm m}$)
can be set to a different value than $n$.
There are six buffer points per direction on each sides
of refinement grid; cf.\ Refs.~\cite{Bruegmann:2006at,Bruegmann:2003aw}
for more information about the buffer points.
For simplicity, we always quote grid sizes without buffer points.
For the wave zone, a shell made up from six ``cubed sphere'' patches
~\cite{Pollney:2009yz,Thornburg2000:multiple-patch-evolution,
Thornburg2004:multipatch-BH-excision} can be added. This helps to improve
the accuracy in GW extraction and allows the implementation
boundary conditions derived for spherical geometries, see
e.g., \cite{Ruiz:2010qj}.

\subsection{Spacetime and Matter Evolution}
\label{sec:method:gravity}
\label{sec:method:matter}
\label{sec:eos}
We employ the Z4c formulation of the Einstein Equations~\cite{Ruiz:2010qj,Hilditch:2012fp,Bernuzzi:2009ex}
combined with the moving puncture gauge using the 1+log-slicing condition~\cite{Bona94b}
and the Gamma driver shift~\cite{Alcubierre02a,vanMeter:2006vi}.
For our single star evolutions, Sommerfeld boundary conditions~\cite{Sommerfeld:1949a} are used.
For binary neutron stars, we add spherical patches outside of
the coarsest cubic box to allow the use of constraint-preserving
boundary conditions~\cite{Hilditch:2012fp}.

We assume that the matter is properly described as a perfect fluid
for which the stress-energy tensor is given by
\be
  T^{\mu \nu} = \left( e + P \right) u^\mu u ^\nu
  + P g^{\mu \nu}, \label{eq:Tmunu}
\ee
with the energy density $e$, the pressure $P$,
and the four-velocity $u^\mu$.
The total energy density is given by
$e=\rho(1+\epsilon)$, where $\rho$ is the rest-mass energy density
and $\epsilon$ is the specific internal energy. In many equations we also
use the specific enthalpy given by
\be
h\equiv1+\epsilon + P/\rho .
\ee

The matter equations follow from the conservation law for the energy-momentum
tensor and the conservation law for the baryon number.
Following~\cite{Banyuls97} the equations governing the evolution of general relativistic fluids,
can be written in first-order flux-conservative form
\bea
  \label{eq:hydro_consform}
  \partial_t \vec{q} + \partial_i \vec{f}^{(i)} (\vec{q}) = \vec{s}(\vec{q})
\ ,
\eea
with $\vec{q}$ denoting the conserved variables, $\vec{f}^{(i)}$ the
fluxes, and $\vec{s}(\vec{q})$ the source terms.
The conserved variables are rest-mass density ($D$),
the momentum density ($S_i$), and internal energy ($\tau$)
as seen by Eulerian observers.
The conserved variables are related to the original variables via
\bea
\label{D-def}
D    &=& \rho W \\
\label{Si-def}
S_i  &=& \rho h W^2 v_i \\
\label{tau-def}
\tau &=& \rho h W^2 - P - \rho W ,
\eea
where $v_i$ is the 3-velocity and $W$ the Lorentz factor of the fluid.


To close the evolution system, we have to specify an
EOS for the fluid.
We choose to employ a simple ideal gas EOS in our single
star evolutions and a piecewise polytropic fit of the zero-temperature SLy
EOS~\cite{Chabanat:1997qh, Douchin:2001sv} for which we add an additional
ideal gas thermal contribution~\cite{Shibata:2005ss}
with $\Gamma^{\rm hot}=1.75$~\cite{Bauswein:2010dn}.

\subsection{Dealing with low density or vacuum regions}

\subsubsection{Original implementation in BAM}
\label{sec:method:atm}

NSs surrounded by vacuum are modeled in numerical
relativity simulations by using an artificial atmosphere,
e.g.,~\cite{Font98b,Dimmelmeier02a,Baiotti04a,Thierfelder:2011yi}.
The artificial atmosphere outside of the stars
is chosen as a fraction of the initial central density of the star as
$\rho_{\rm atm}\equiv f_{\rm atm}\cdot \rho_c(t=0)$.  The atmosphere
pressure and internal energy is computed by employing the
zero-temperature part of the EOS. The fluid velocity within the
atmosphere is set to zero.  At the start of the simulation, the
atmosphere is added before the first evolution step.  During the
recovery of the primitive variables from the conservative variables, a
point is set to atmosphere if the density is below the threshold
$\rho_{\rm thr}\equiv f_{\rm thr}\rho_{\rm atm}$.  
In this article, we are using $f_{\rm atm} = 10^{-11}$ 
and $f_{\rm thr} = 10^2$ in all test cases.

\subsubsection{A new vacuum treatment}
\label{sec:method:vacuum}

\textit{\textbf{Conservative to primitive conversion:}}
Unfortunately, in general there is no closed analytic expression for the primitive
variables in terms of the conserved ones. We thus have to resort to a root
finder. Within our new vacuum treatment, we use the following
scheme. We square Eq.~(\ref{Si-def}) and use the definition of the
conservative variables to find
\be
W^2 = \frac{(D + \tau + P^*)^2}{(D + \tau + P^*)^2 - S^2} .
\ee
Here, $P^*$ is an initial guess for the pressure and we have
defined $S^2 = \gamma^{ij}S_i S_j$.
Once $W(P^*)$ is known, we can solve Eqs.~(\ref{D-def}) and (\ref{tau-def})
for $\rho$ and $\epsilon$. We obtain
\be
\rho(P^*) = \frac{D}{W(P^*)}
\ee
and
\be
\label{epsilon-of-P}
\epsilon(P^*) = \frac{\sqrt{(D + \tau + P^*)^2 - S^2} - W(P^*) P^*}{D} - 1 .
\ee
Using a one dimensional root finder, we
adjust $P^*$ until the EOS of the form $P = P(\rho, \epsilon)$ is satisfied.
However, both $W(P^*)$ and $\epsilon(P^*)$ contain a square root
of $(D + \tau + P^*)^2 - S^2$. Thus we need $P^* > S - D - \tau$.
Furthermore we expect the pressure to be positive. Thus, we need a root
finder that searches for the root $P^*$ in the interval $[P_{min},\infty)$,
where
\be
P_{min} = \min (0, S - D - \tau) .
\ee
Our algorithm employs a Newton-Raphson scheme, but falls back on
bisection whenever the Newton step would bring us outside the allowed
interval. In addition, we limit Eq.(\ref{epsilon-of-P}) to not violate the
weak energy condition, i.e. whenever $\epsilon(P^*) \leq -1$, we
set it to $-(1.-10^{-10})$. In most cases we can then find a root and obtain
a suitable $P^*$. In case where this is not possible,
we reset all variables to vacuum, i.e. we set
\be
\label{cons_prim_vacuum}
D = \tau = S_i = \rho = \epsilon = P = v_i = 0 .
\ee
We also reset all variables to vacuum if we find that $D<0$, since negative
rest mass densities are non-physical.
We point out that similar checks are also present with artificial atmospheres,
but with larger threshold values. \\

\textit{\textbf{Reconstruction and fluxes:}}
The evolution equations for the conserved fluid variables are computed from
fluxes at cell interfaces where we do not have grid points. In order to
obtain these fluxes, we interpolate the quantities necessary, to compute them
at the cell interface locations. For the smooth gravitational fields such as
$\alpha$, $\beta^i$, and $\gamma_{ij}$ we use sixth order Lagrangian
interpolation, while for the potentially non-smooth matter fields we use 5th
order WENOZ interpolation for the primitive variables~\cite{Borges:2008a,Bernuzzi:2016pie}.
The interpolation results at each interface are constructed in two ways:
once from data to the
left (L) of the interface and again from data to the right (R) of the
interface. For the primitive variables this results in $\rho_{L/R}$,
$\epsilon_{L/R}$ and $Wv_{i\ L/R}$\footnote{Notice that we interpolate
$Wv_{i}$ and not $v_{i}$ to avoid cases where the 3-velocity is interpolated
to a value above light speed.}. Interpolation can still lead to unphysical
values on either side. If the determinant of $\gamma_{ij}$ is less than or
equal to zero, we set it to 1 and also set $\rho_{L/R} = \epsilon_{L/R} = 0$.
Furthermore,  if $\rho_{L} < 0$ we set $\rho_{L} = \epsilon_{L} = 0$ and
if $\rho_{R} < 0$ we set $\rho_{R} = \epsilon_{R} = 0$.
In order to obtain the pressure $P_{L/R}$ as well as the sound speed squared
$
c^2_{s\ L/R} = \frac{1}{h}
               \left(\frac{\partial P}{\partial\rho} +
                     \frac{P}{\rho^2}\frac{\partial P}{\partial\epsilon}
               \right)_{L/R}
$
we use the EOS.
If $c^2_{s}<0$  or $c^2_{s}>1$
we set it to zero, we also set it to zero if $\rho=0$ or $h=0$. We use the
thus interpolated and limited primitive variables to compute the conserved
variables as well as the fluxes $\vec{f}_{L/R}$ at both interfaces.
In addition, we compute the speeds $\vec{\lambda}_{L/R}$ of the
characteristic variables on both sides using
\bea
\lambda_1 &=& \alpha \frac{v^n (1-c^2_s) + \sqrt{C^2}}{1-v^2 c^2_s} - \beta^n \\
\lambda_2 &=& \alpha \frac{v^n (1-c^2_s) - \sqrt{C^2}}{1-v^2 c^2_s} - \beta^n \\
\lambda_3 &=& \alpha v^n - \beta^n \\
\lambda_4 &=& \alpha v^n - \beta^n \\
\lambda_5 &=& \alpha v^n - \beta^n
\eea
where
$C^2 = c^2_s\{ (1-v^2)[ g^{nn}(1-v^2 c^2_s) - v^n v^n(1-c^2_s) ] \}$,
$v^n = v^i n_i$ and $n_i$ is the normal to the interface.
If $1-v^2 c^2_s = 0$ or $C^2<0$ we simply set $\lambda_1=\lambda_2=0$.

The final numerical flux
$\vec{F}$ at the interface is then computed using a standard method
such as the Local Lax-Friedrichs (LLF) scheme where
\be
\vec{F}_{interface}
=\frac{1}{2}[\vec{f}_R + \vec{f}_L - |\lambda|_{max}(\vec{q}_R - \vec{q}_L)]
.
\ee
Here $\vec{q}_{L/R}$ are the conserved variables on the left or right and
$|\lambda|_{max}$ is the characteristic speed with the largest magnitude. In
fact, in the simulations presented in this paper we use the LLF flux
always at low densities, while possibly using a higher order flux at higher
densities. In this case, the higher-order flux $\vec{F}_{interface}^{HO}$ is
obtained by interpolating the characteristic variables from five neighboring
points using the WENOZ scheme~\cite{Bernuzzi:2016pie}. In some simulations
labeled with HO we use this higher order flux $\vec{F}_{interface}^{HO}$
above a certain density threshold (typically on the order of $1\%$ of the
maximum of $\rho$ at the star center).\\

\textit{\textbf{Matter removal:}} We use a forth order Runge-Kutta
scheme to evolve the conserved variables.  Before we evaluate the
right hand side within each Runge-Kutta substep, we set the conserved
variables to vacuum if one of the following two conditions is true:
(1) if $D < 0$, (2) if $D > f W \rho$ and $\alpha < 0.2$, where the
factor $f$ is usually chosen to be $100$.  The first condition is
obvious and allows only positive matter density. The second case is
used for matter removal inside black holes (BHs). Since we use the standard
moving puncture gauges 1+log-lapse and gamma-driver
shift~\cite{Bona94a,Alcubierre02a,vanMeter:2006vi}, the BH
horizon is located near the surface where $\alpha \sim 0.3$.
Therefore, condition (2) is true only inside the horizon. We have
observed that when matter accumulates near the BH center after
a star collapses, $D$ rises much faster than $\rho$ so that
Eq.~(\ref{D-def}) is violated. This happens because the BH
center in this gauge is only very poorly
resolved~\cite{Hannam:2008sg,Dietrich:2014cea}.  Condition (2) ensures
that matter is removed whenever $D$ becomes much larger than
$\rho$.
An alternative approach was described in
\cite{Thierfelder:2010dv}, where matter is not removed, but some of
the eigenvalues and $W$ are reset for regularity, which could be
explored in future work.

\section{Single Star Spacetimes}
\label{sec:single-star}

\begin{table}[t]
  \centering
  \caption{Parameters and properties of the single neutron star tests.
          We report the density at which we switch between primitive and characteristic
          reconstruction ($\rho_{\rm switch}$),
          threshold density and atmosphere density for the artificial
          atmosphere ($\rho_{\rm thr}$ \& $\rho_{\rm atm}$),
          EOS parameters $\Gamma$ and $K$ for the polytropic EOS
          ($p=K \rho^\Gamma$) to setup the initial data, and
          employed symmetry to reduce computational costs.
          For completeness, we also present the gravitational mass $M$,
          the baryonic rest-mass $M_b$, the initial central density $\rho_c$,
          the equatorial radius $R_e$, and the aspect ratio $R_p/R_e$.}
  \begin{tabular}{|l|c|c|c|c|}
    \hline
    Stars           & TOV$_{static}$ & TOV$_{mig}$    & RNS$_{col}$ \\
    \hline
    $\rho_{switch}\cdot 10^{-5}$ & $1.28$  & $7.83$   & $3.12$ \\
    $\rho_{thr}\cdot 10^{-12}$   & $1.280$ & $7.9934$ & $3.1160$ \\
    $\rho_{atm}\cdot 10^{-14}$   & $1.280$ & $7.9934$ & $3.1160$ \\
    $\Gamma$                     & 2       & 2        & 2 \\
    $K$                          & 100     & 100      & 99.5 \\
    Symmetry                     & octant  & octant   & quadrant \\
    \hline
    $M$                          & 1.400   & 1.448    & 1.861 \\
    $M_b$                        & 1.506   & 1.535    & 2.044 \\
    $\rho_{c}\cdot 10^{-3}$      & 1.2800  & 7.9934   & 3.1160 \\
    $R_e$                        & 8.126   & 4.268    & 9.652 \\
    $R_p/R_e$                    & 1       & 1        & 0.65 \\
    \hline
  \end{tabular}

 \label{Tab:tests-runs}
\end{table}

\begin{table}[t]
  \centering
  \caption{The grid parameters for the single star configurations
    at all four resolutions are tabulated here.
    The atmosphere and vacuum treatments have the same parameters and thus are
    tabulated only once.
    $L$ is the total number of boxes,
    $n$ $(n^{mv})$ is the number of points in the fixed (moving) boxes, and
    $h_{0},h_{L-1}$ are the grid spacing in level $l=0,L-1$.
    The grid spacing in level $l$ is $h_l=h_0/2^l$.}
  \begin{tabular}{|c|cccccc|}
    \hline
     & & \multicolumn{5}{c|}{Grid Parameters} \\
     \hline
     Tests & Resolutions  & $L$ & $n$ & $n^{mv}$ & $h_0$ & $h_{L-1}$ \\
    \hline
     \multirow{4}{*}{TOV$_{\rm static}$} & Low(L) & 5 & 64 & 64 & 1.125 & 0.281 \\
      & Med(M) & 5 & 96 & 96 & 0.750 & 0.188 \\
      & High(H) & 5 & 128 & 128 & 0.563 & 0.141 \\
      & Fine(F) & 5 & 160 & 160 & 0.450 & 0.113 \\
     \hline
     \multirow{4}{*}{TOV$_{\rm mig}$} & Low(L) & 7 & 64 & 64 & 19.20 & 0.300 \\
      & Med(M) & 7 & 96 & 96 & 12.80 & 0.200 \\
      & High(H) & 7 & 128 & 128 & 9.600 & 0.150 \\
      & Fine(F) & 7 & 160 & 160 & 7.680 & 0.120 \\
     \hline
     \multirow{4}{*}{RNS$_{\rm col}$} & Low(L) & 9 & 64 & 28 & 18.00 & 0.070 \\
      & Med(M) & 9  & 96 & 42 & 12.00 & 0.047 \\
      & High(H) & 9 & 128 & 56 & 9.00 & 0.035 \\
      & Fine(F) & 9 & 160 & 72 & 7.20 & 0.028 \\
     \hline
  \end{tabular}
 \label{Tab:tests-grid1}
\end{table}

To test our new implementation, we start by studying three different single star configurations:
\begin{enumerate}
 \item a stationary, static star (TOV$_{\rm static}$ -- Sec.~\ref{sec:tests:TOV})
 \item an unstable, migrating star (TOV$_{\rm mig}$ --  Sec.~\ref{sec:tests:migration})
 \item a perturbed, collapsing, uniformly rotating neutron star
  (RNS$_{\rm col}$ -- Sec.~\ref{sec:tests:Rcollapse})
\end{enumerate}

Each test uses the LLF flux with primitive
reconstruction~\cite{Kurganov:2000, LucasSerrano:2004aq,Nessyahu:1990}, labeled as LLF,
and the hybrid scheme employing characteristic reconstruction for large
and primitive reconstruction for low densities (labeled as HOLLF).
We employ four different resolutions denoted as $\rm Low$, $\rm Mid$, $\rm High$, and $\rm Fine$.
Details about the physical setup and the grid parameters are given in Tab.~\ref{Tab:tests-runs}
and Tab.~\ref{Tab:tests-grid1}, respectively.
In order to assess the performance of the old `atmosphere' and the new
`vacuum' method, we compare the central density, the total rest-mass,
and the Hamiltonian Constraint during the evolution for all tests.

\subsection{Stationary TOV simulations}
\label{sec:tests:TOV}

\begin{figure*}[t]
\includegraphics[width=1.\textwidth]{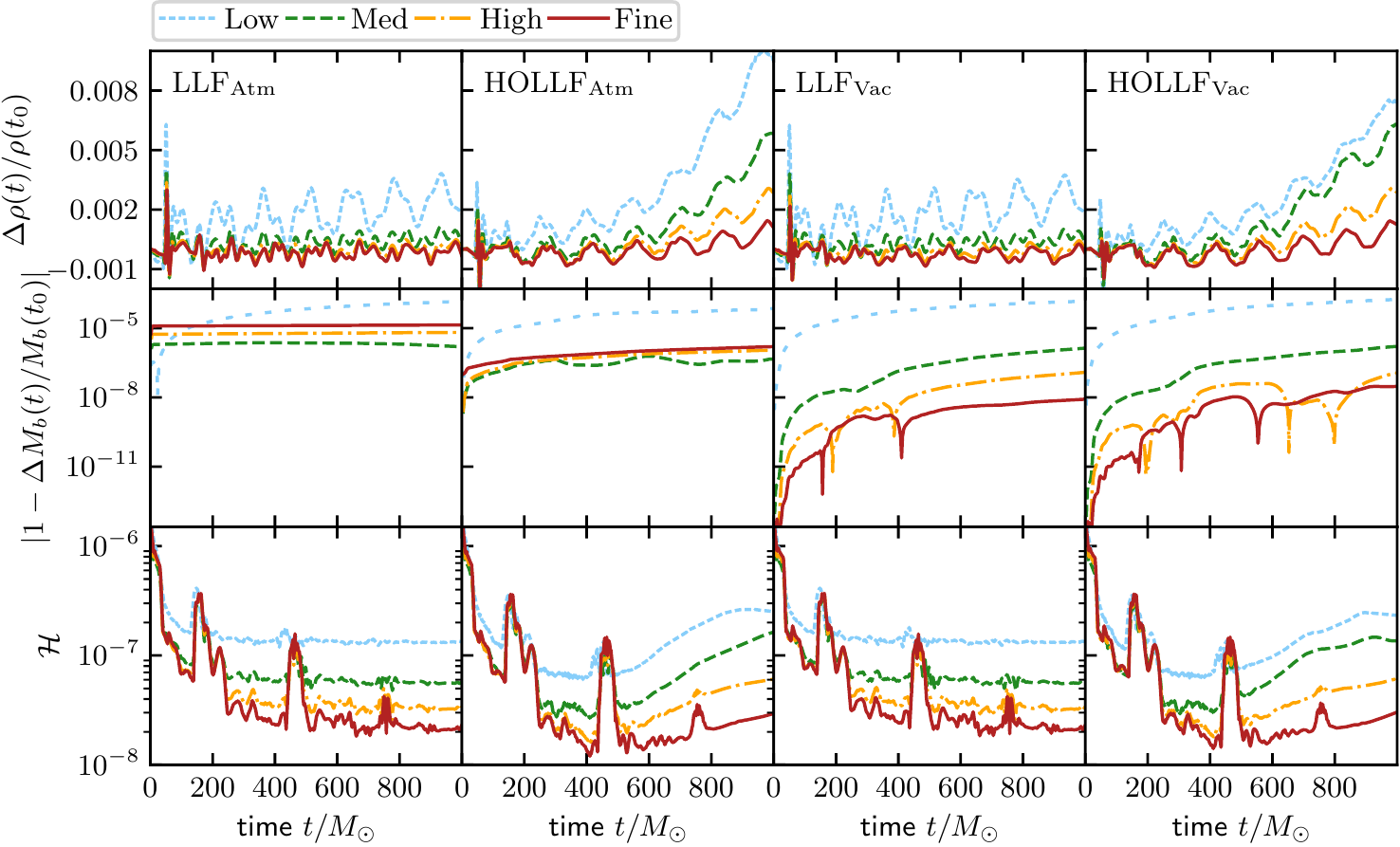}
\caption{Results of the TOV$_{static}$ test.
  Left to right: Atmosphere-LLF, Atmosphere-HOLLF, Vacuum-LLF, and Vacuum-HOLLF.
  Top: Relative change in central density $1 - \frac{\rho_c(t)}{\rho_c(t=0)}$.
  Middle: Relative rest-mass change $|1 - \frac{M_b(t)}{M_b(t=0)}|$.
  Bottom: The time evolution of Hamiltonian Constraint($\mathcal{H}$).}
\label{fig:test-static}
\end{figure*}

In Fig.~\ref{fig:test-static}, we plot the relative central
density $1 - \frac{\rho_c(t)}{\rho_c(t=0)}$, the relative rest-mass change
  $|1 - \frac{M_b(t)}{M_b(t=0)}|$, and the Hamiltonian constraint for all
TOV$_{\rm static}$ simulations. All quantities are extracted at level $l=4$
which is the finest level, but also fully covers the entire star.
The stars are evolved up to a time of 1000$M_{\odot}$, i.e., $4.93\rm ms$.
Truncation errors trigger small-amplitude pulsations in
the stars~\cite{Font98b, Font01b} that can be seen as oscillations
in the relative central density.
The central density oscillations are larger for the hybrid HOLLF method, but
decrease clearly with an increasing resolution. There is no noticeable difference
between the old atmosphere and the new vacuum method.

Considering the mass conservation, one sees a clear advantage of our
new implementation.  In the case of the old atmosphere method, the
limit for setting the density to the atmosphere value is 1.28 $\cdot
10^{-12}$. In the first few time steps, the star surface slightly
grows causing the density to drop below this threshold. This leads to a
visible violation of mass conservation after the first timestep.
With our new vacuum approach a low density layer builds up around the
star. Thus with our new approach mass is much better conserved.
In addition, even during the subsequent evolution one observes a larger mass
violation for the atmosphere method than for our new implementation,
where for the highest resolution the mass violation is below $10^{-8}$.

It is also important to point out that for the atmosphere case,
we do not observe convergence in the mass. This applies to both the HOLLF
and the LLF scheme.
On the contrary, for the new vacuum method we find second
order convergence for most of the time for LLF and up
to $t=600M_\odot$ for HOLLF.

\begin{figure}[t]
\includegraphics[width=0.5\textwidth]{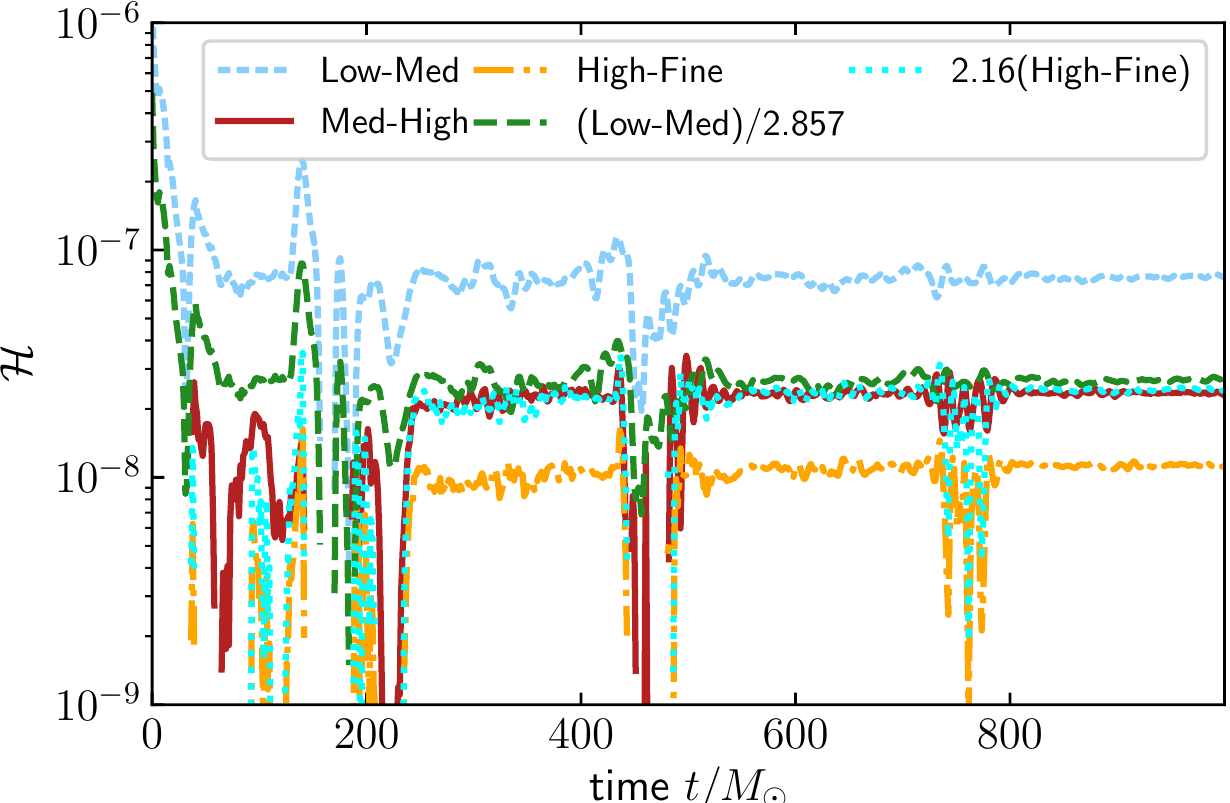}
\caption{
Convergence test of Hamiltonian Constraint of LLF Vacuum case of
TOV$_{static}$ neutron star. The dotted cyan and dashed green lines are
obtained by scaling the dash-dotted orange line and the dashed blue line
respectively in order to match up with the solid red line.
}
\label{fig:convergence}
\end{figure}

The bottom panel of Fig.~\ref{fig:test-static} shows the evolution of the
Hamiltonian constraint. We observe a reduction of the Hamiltonian constraint
for increasing resolution exhibiting clean second-order convergence.
As an example we show a convergence test for the LLF vacuum setup in
Fig.~\ref{fig:convergence}. Here
the differences in the Hamiltonian Constraint are scaled by factors
that correspond to assuming second order convergence.
These scaled lines nicely coincide
with the middle line as expected for second order convergence.

\subsection{Migration of an unstable star}
\label{sec:tests:migration}

\begin{figure*}[t]
\includegraphics[width=\textwidth]{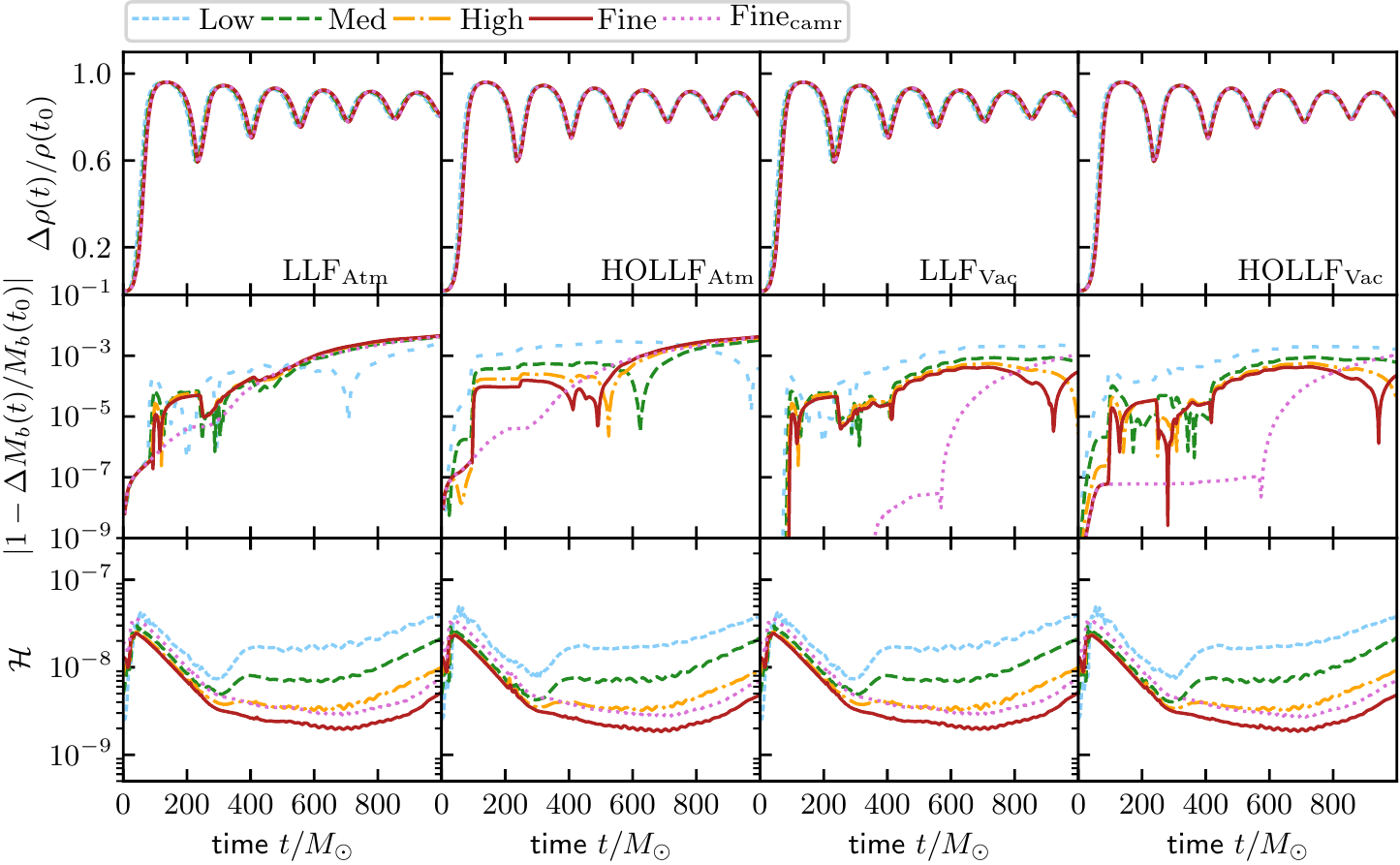}
\caption{Results of the TOV$_{\rm mig}$ test.
  Left to right: Atmosphere-LLF, Atmosphere-HOLLF, Vacuum-LLF, and Vacuum-HOLLF.
  Top: Relative change in central density $1 - \frac{\rho_c(t)}{\rho_c(t=0)}$.
  Middle: Relative rest-mass change $|1 - \frac{M_b(t)}{M_b(t=0)}|$.
  Bottom: The time evolution of Hamiltonian Constraint($\mathcal{H}$).
  Since matter is expected to cross refinement boundaries during this test,
  we also perform for the highest resolution, a simulation in which we
  apply the conservative refluxing algorithm that we
  developed in~\cite{Dietrich:2015iva}.}
\label{fig:test-mig}
\end{figure*}
\begin{figure*}[t]
\includegraphics[width=0.45\textwidth]{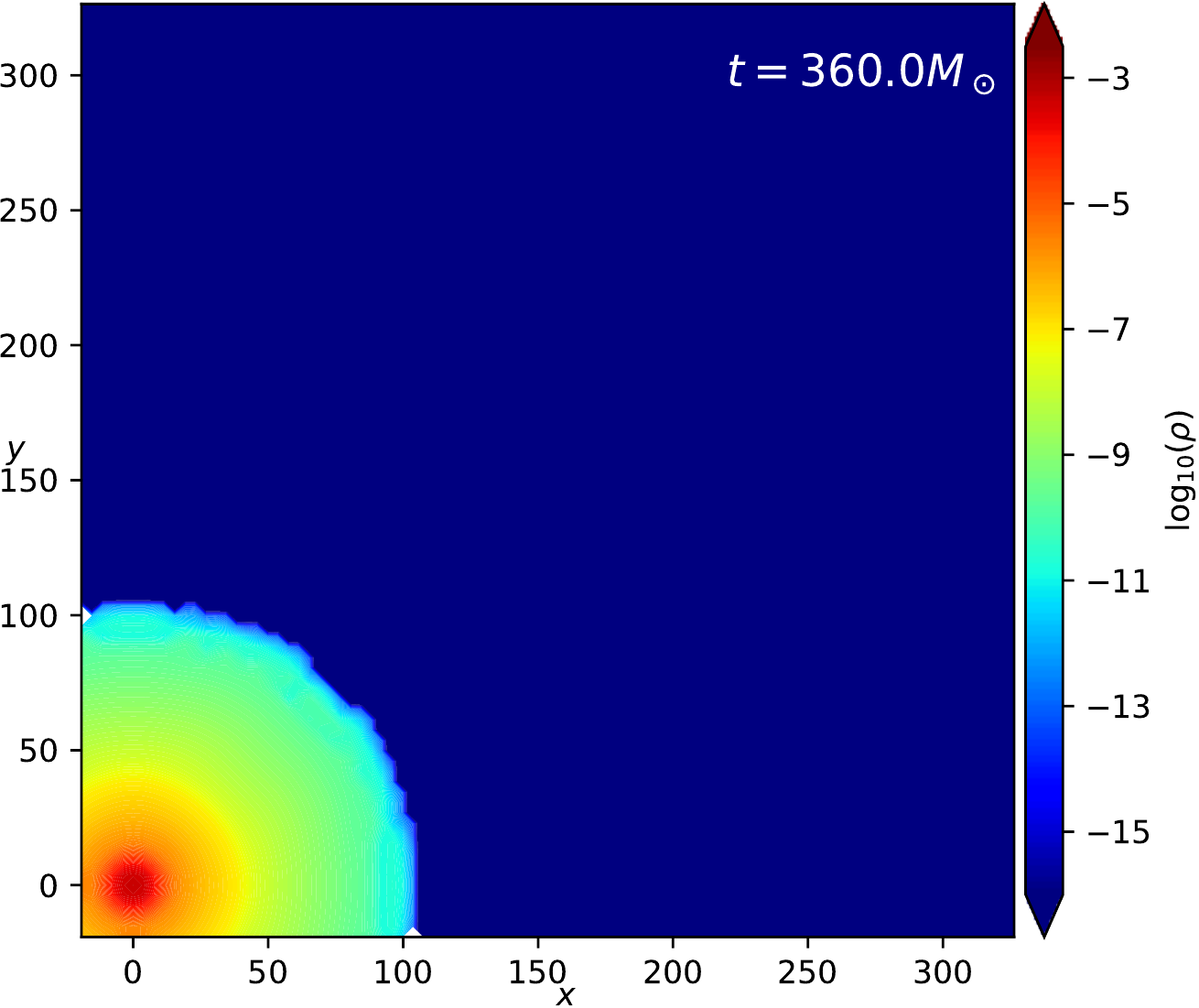} \hfill
\includegraphics[width=0.45\textwidth]{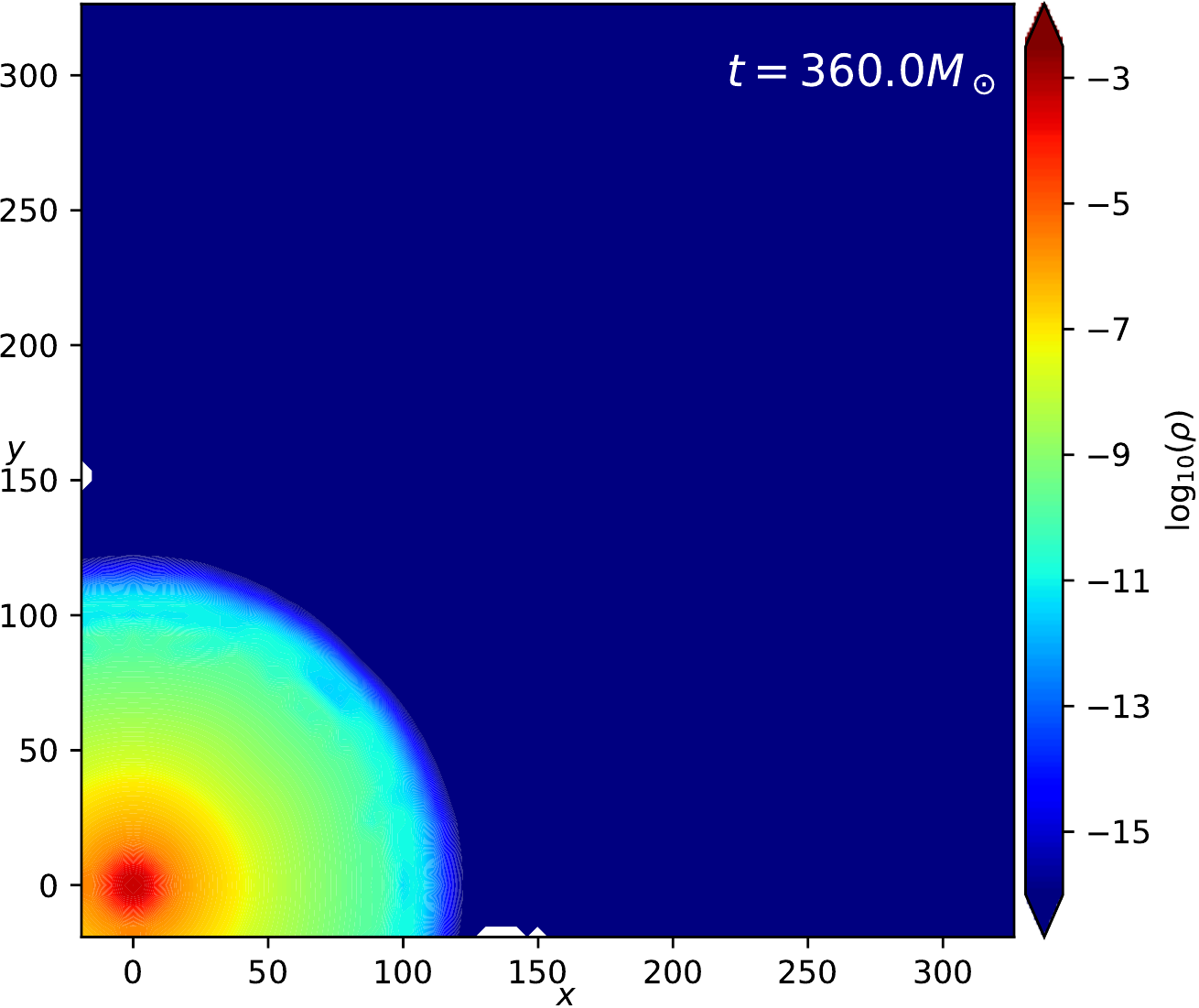} \\
\includegraphics[width=0.45\textwidth]{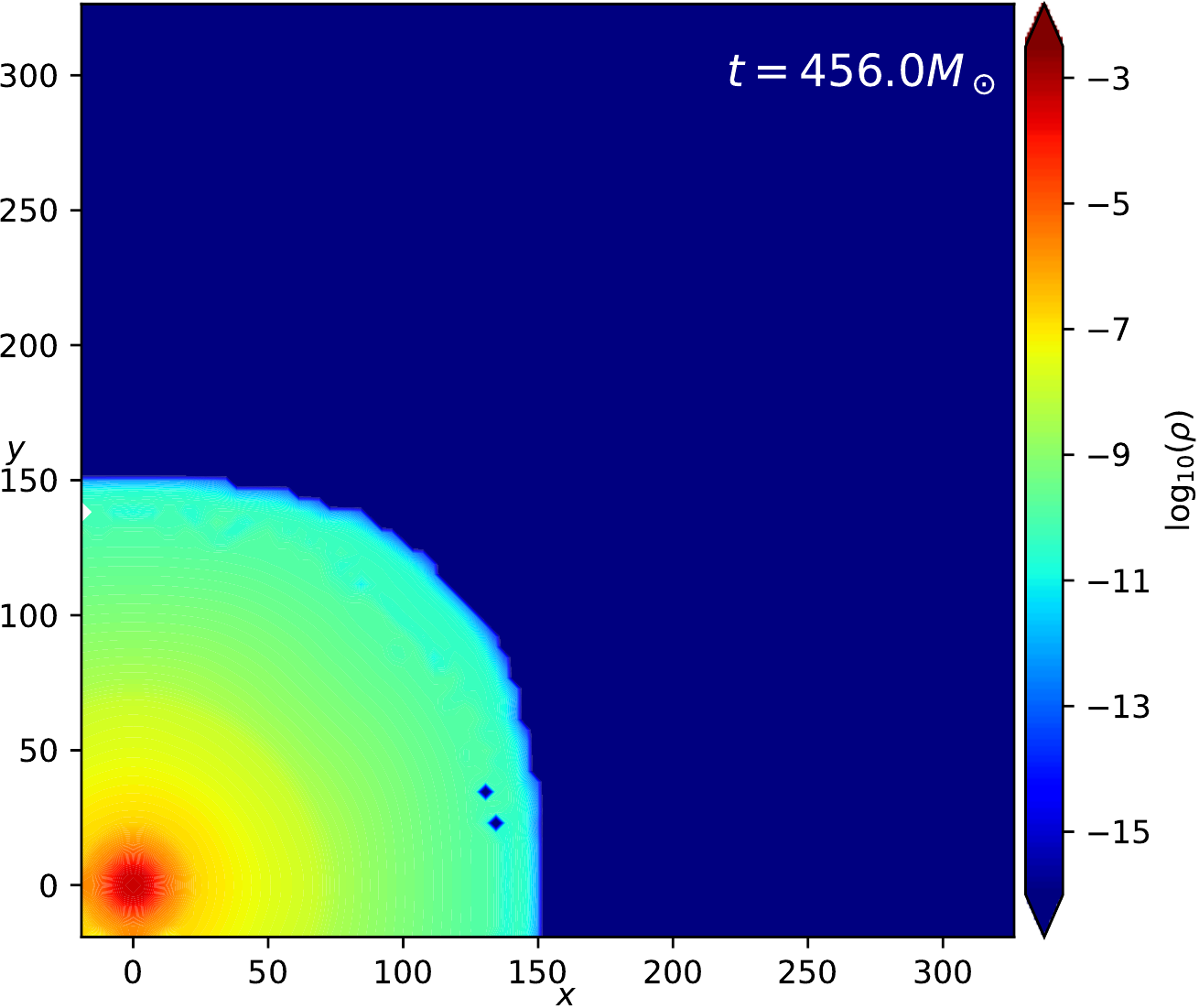} \hfill
\includegraphics[width=0.45\textwidth]{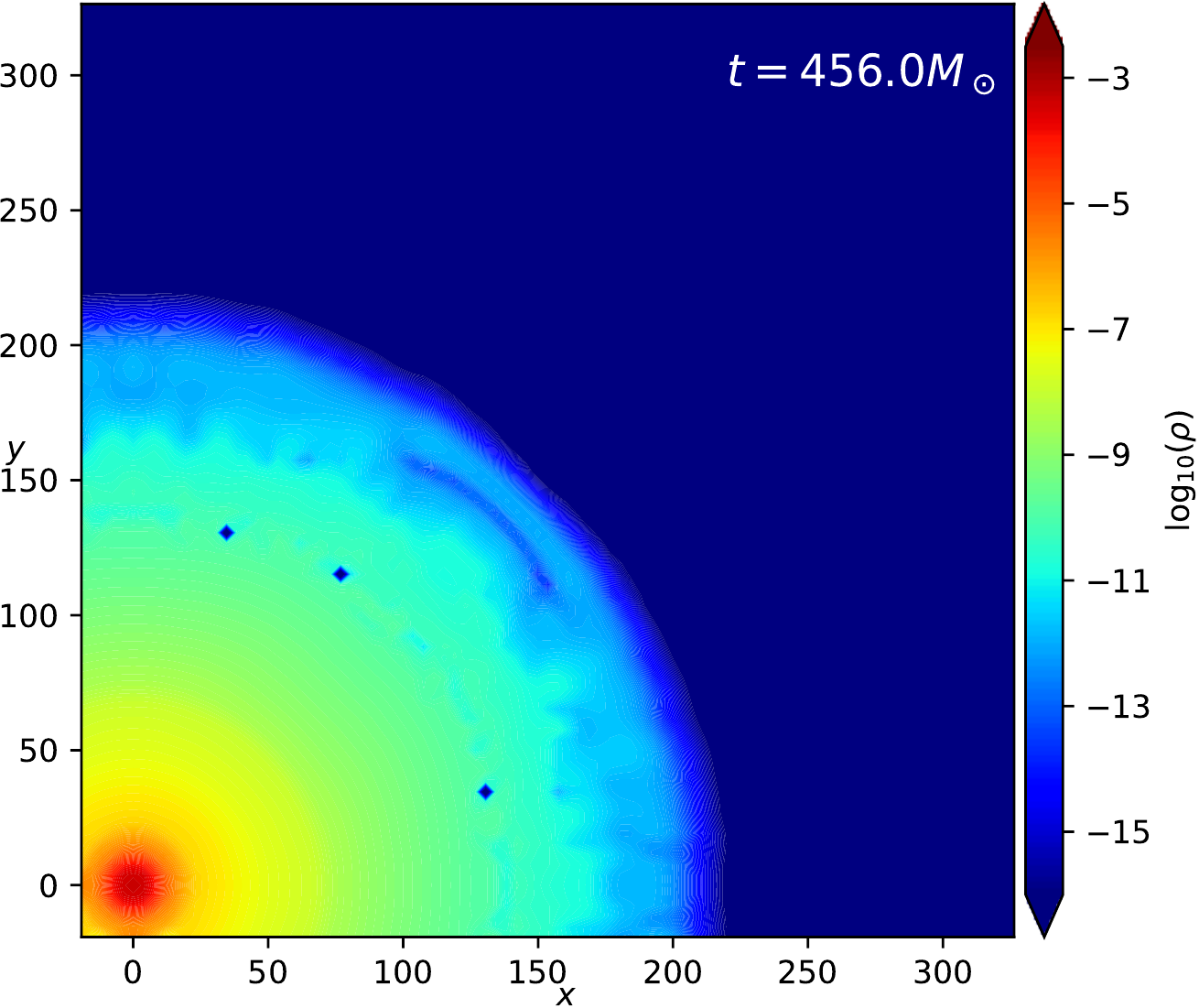} \\
\includegraphics[width=0.45\textwidth]{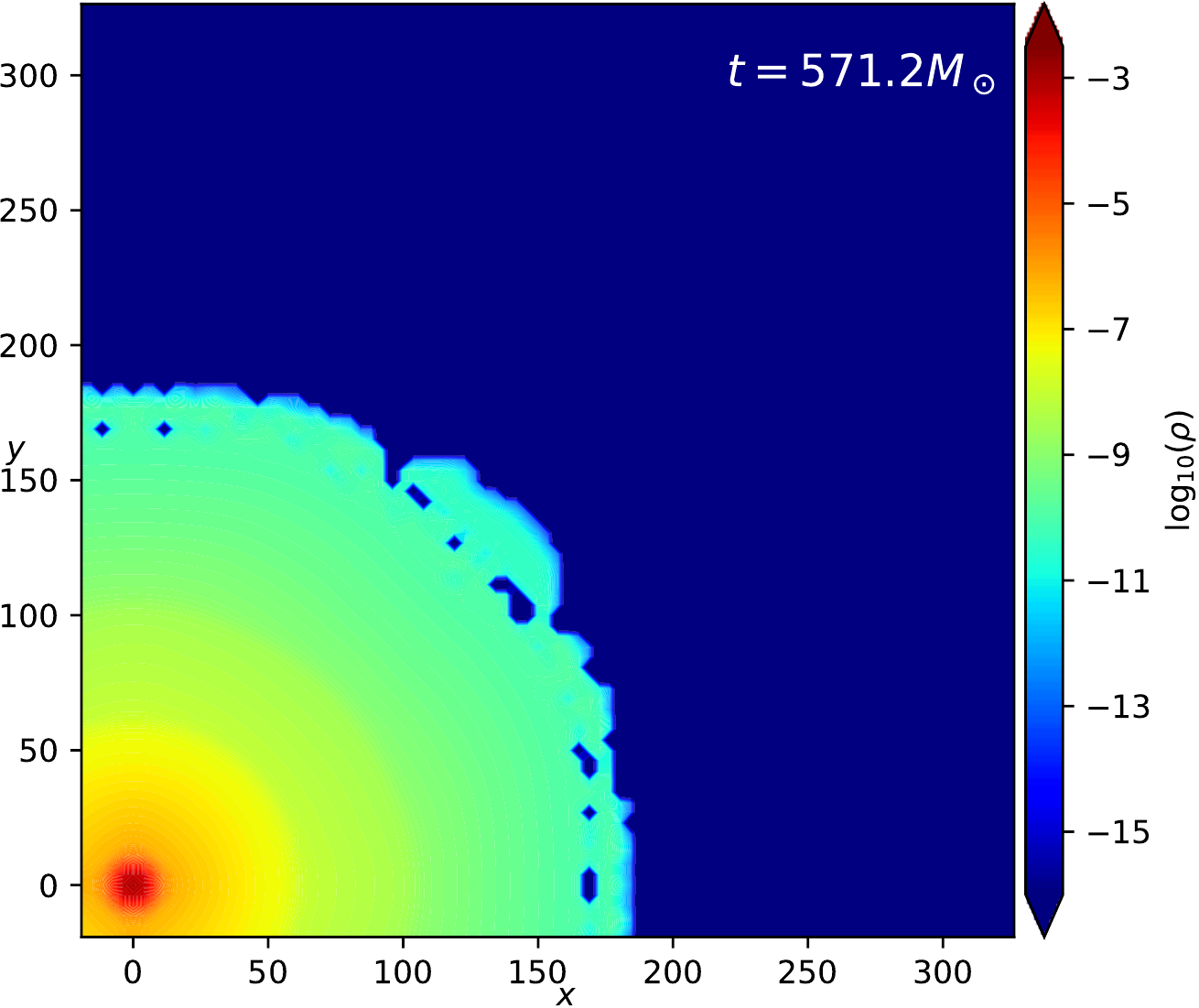} \hfill
\includegraphics[width=0.45\textwidth]{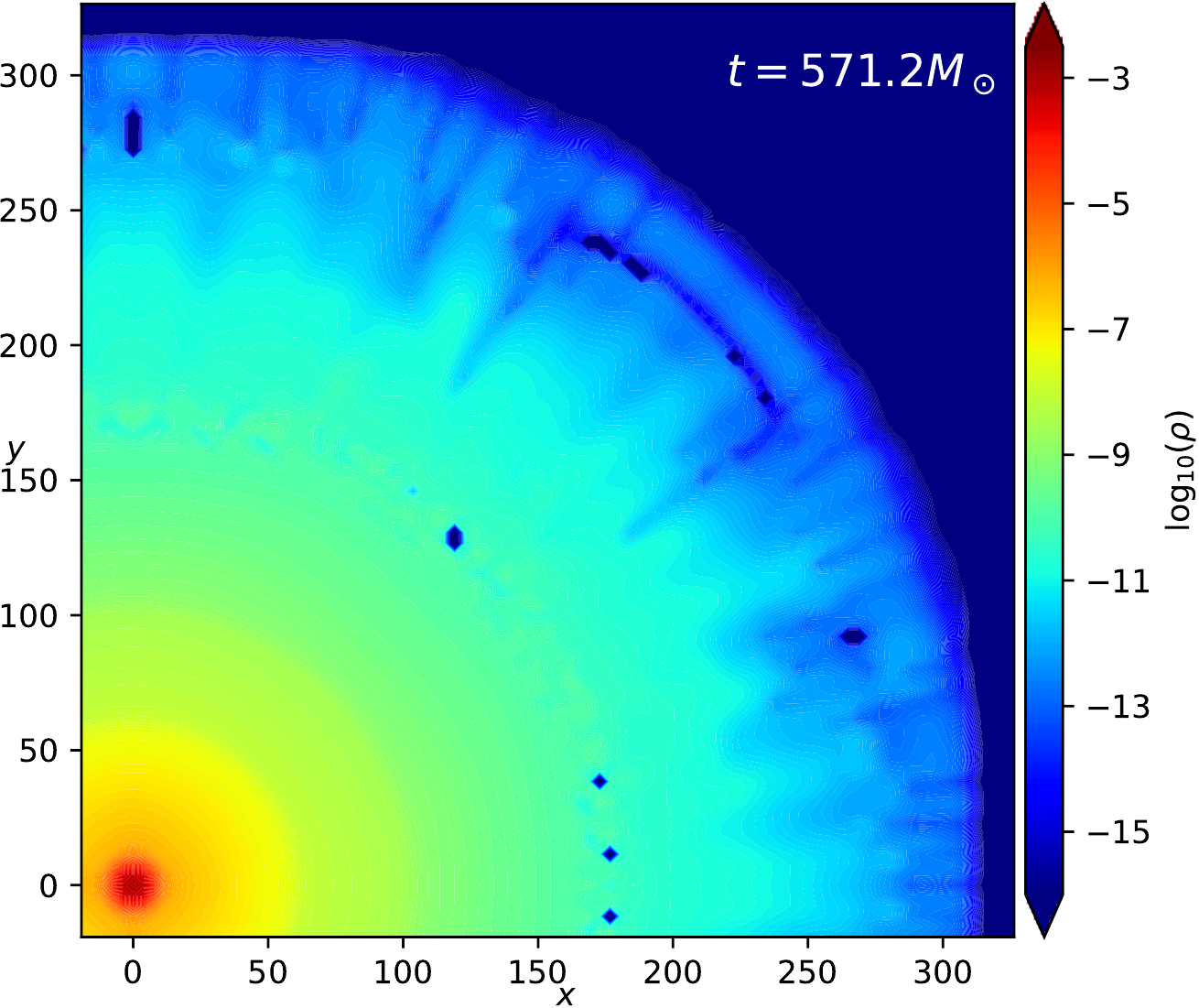} \\
\caption{The rest-mass density in xy-plane for the TOV$_{\rm mig}$ test on refinement level
$l =1$ at different times for the finest resolution (n=160 points) employing the conservative
mesh refinement~\cite{Dietrich:2015iva}.
The left panels show the previous atmosphere scheme, the right panel shows the new vacuum implementation.
For the atmosphere case, the threshold density below which artificial atmosphere is
set up $\rho_{thr}$ is $\sim 7.9934\cdot10^{-12}$ and the artificial
atmosphere level($\rho_{atm}$) is $\sim 7.993\cdot10^{-14}$.}
\label{fig:mig-2D}
\end{figure*}

TOV$_{\rm mig}$ represents a test based on the unstable U0 model of Ref.~\cite{Thierfelder:2011yi}, with a central energy density of
$\epsilon_c=8.73\times10^{-3}$ and a gravitational mass of $M=1.557$,
A small perturbation caused by truncation errors leads to pulsations
that migrate the star towards a stable configuration of the same rest-mass.
Initially, the central density decreases and the star expands rapidly.
Later its inner core contracts which leads to a shrinking of the star
and an increase of the central density. As a result, it pulsates
causing matter to cross the grid refinement boundaries.
To better resolve the dynamics we are using a larger number of refinement levels than in
the TOV$_{\rm static}$ test.

In Fig.~\ref{fig:test-mig} we plot the central density on
the finest level $l=6$ and the rest-mass and
Hamiltonian constraint on level $l=1$.
In the top panels, we see a decrease in the amplitude of pulsation of
central density as the simulation progresses. If we would run the simulation
longer, the star would finally settle down to a stable configuration.
The Hamiltonian constraint in the bottom
panel converges roughly with a second-order in all four cases. For the LLF
case we see convergence throughout the simulation whereas in HOLLF we see
convergence roughly from 300$M_{\odot}$ to 1000$M_{\odot}$.
Thus considering the Hamiltonian constraint,
the LLF simulations perform better than the HOLLF ones.

As for TOV$_{\rm static}$ case, we find a better mass conservation
for the vacuum configurations than for the old atmosphere method.
Convergence consistent with the second-order is observed in the
early part of the simulations HOLLF simulations,
for the LLF method no convergence is present at all.
During the evolution time and because of the pulsation of the star,
mass is crossing the refinement level.
At this time, mass conservation is generally lost if no additional
conservative refluxing step as introduced in~\cite{Berger89,Dietrich:2015iva} is
applied. To prove this point, we perform a simulation with the highest
resolution and activate the refluxing scheme, labeled as
$\rm Fine_{\rm camr}$ in Fig.~\ref{fig:test-mig}.
We find that for the Vacuum method mass conservation is significantly
improved up to about $t=600M_\odot$. At this time low density material
hits the outer boundary of the considered computational domain and
leaves it, consequently the total mass can not be conserved after this point.

Considering 2d-snapshots of the matter evolution clearly reveals the
advantage of the new vacuum treatment. As can be seen in
Fig.~\ref{fig:mig-2D}, when very low density material expands, it is stopped
due to the artificial atmosphere (see bottom left panel) while it 
expands freely in the vacuum case (bottom right panel).
Such an artificial impact on the outgoing matter could be of significant
importance if one wants to track outward going ejecta.

\subsection{Collapsing, rotating neutron star}
\label{sec:tests:Rcollapse}
\begin{figure*}[t]
\includegraphics[width=1\textwidth]{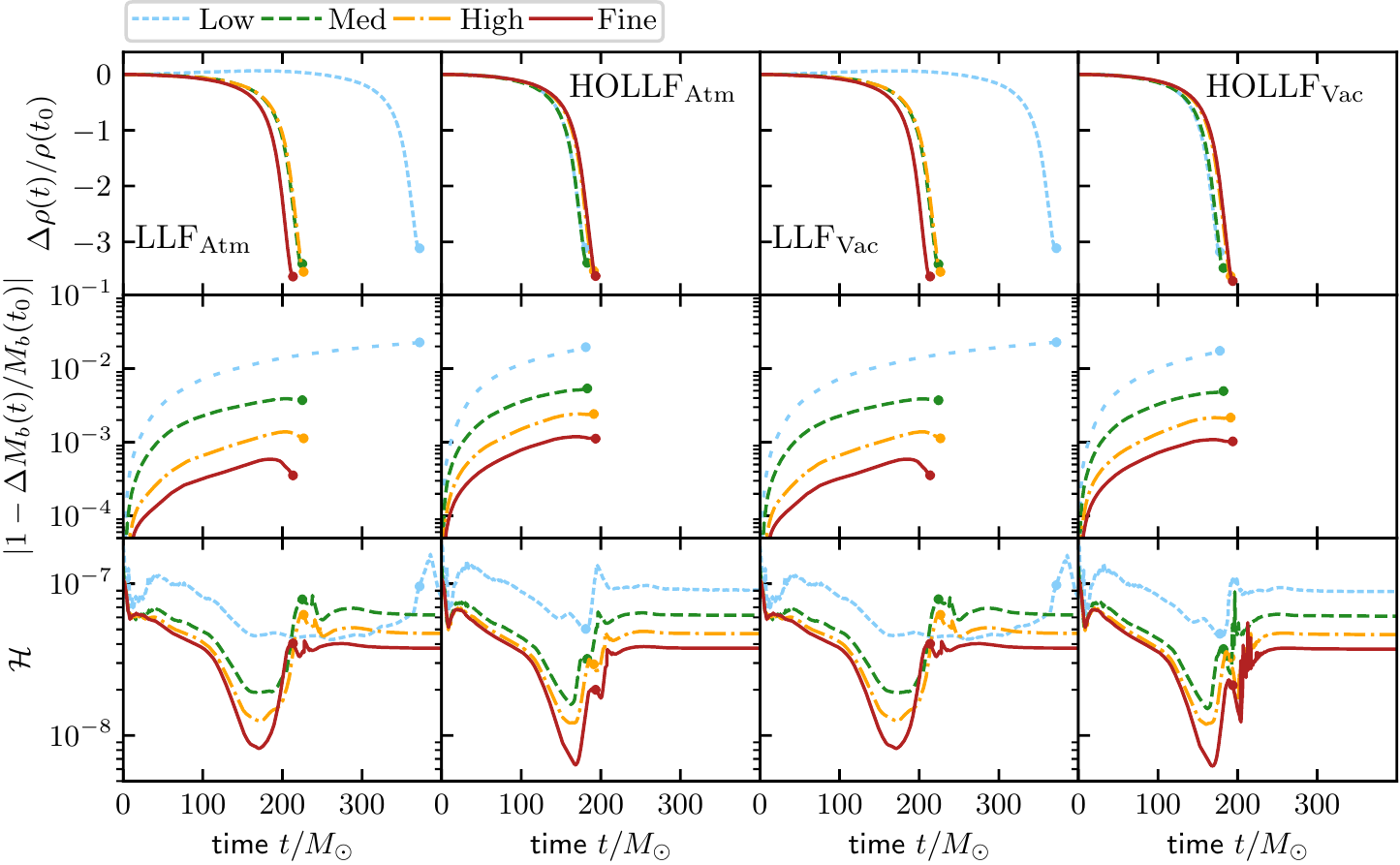}
\caption{Results of the RNS$_{\rm col}$ test.
  Left to right: Atmosphere-LLF, Atmosphere-HOLLF, Vacuum-LLF, and Vacuum-HOLLF.
  Top: Relative change in central density $1 - \frac{\rho_c(t)}{\rho_c(t=0)}$.
  Middle: Relative rest-mass change $|1 - \frac{M_b(t)}{M_b(t=0)}|$.
  Bottom: The time evolution of Hamiltonian Constraint($\mathcal{H}$).}
\label{fig:test-colR}
\end{figure*}
As a last single-star test case,
we study the collapse of a rotating neutron star (RNS).
This test aims towards a better understanding if a BH
can be properly modeled within our new algorithm.

The initial data are computed using a polytropic EOS with
$K=100$, $\rho_c= 3.1160 \cdot 10^{-3}$, and axes ratio $0.65$,
which leads to a star with gravitational mass of $M=1.861$,
baryonic mass of $M_b=2.044$, and an angular velocity of $\Omega=3.96\times10^{-2}$.
The star is evolved with the polytropic EOS with $K=99.5$ and
$\Gamma=2$. This initial perturbation due to the change of the EOS
triggers the collapse of the star to a BH.
A similar configuration has been investigated in
the past, e.g., Refs.~\cite{Giacomazzo:2012bw,Reisswig2013,Dietrich:2014wja}.
We are evolving the star with quadrant symmetry, i.e., use reflection symmetry along the
$x$- and $y$-axis and employ nine refinement levels.

In Fig.~\ref{fig:test-colR} we plot the central density in
the finest level $l=8$ and rest-mass and Hamiltonian constraint
on level $l=3$. The collapse to a BH happens at around $t = 200M_\odot$
for most cases except for the lowest resolution using the LLF scheme.
For both vacuum and atmosphere cases with LLF at $n=64$ points collapse
happens at around $t = 380 M_\odot$. After the star collapses into
a BH, matter is removed to avoid the occurrence of
steep density gradients as mentioned before.

The Hamiltonian constraint shows second-order convergence before the BH formation.
After the collapse, the convergence order reduces to first order.
In both vacuum and atmosphere cases, the error of
rest-mass behaves in a similar way.

Overall, we find no clear and noticeable difference between the old
atmosphere and new vacuum method.

\subsection{Summary of the single star simulations}
We have studied evolutions with an updated implementation of our vacuum treatment
for a number of single star spacetimes.
The main observations are:
\begin{itemize}
 \item Mass conservation can be improved with the new implementation; cf.\ TOV$_{\rm static}$.
 \item The new implementation improves the simulation of outflowing, low density material; cf.\ TOV$_{\rm mig}$;
 \item The new vacuum method is capable of tracking the BH formation; cf.\ RNS$_{col}$.
\end{itemize}

\section{Binary neutron star evolutions}
\label{sec:BNS}

\begin{table}[t]
  \centering
  \caption{The grid parameters for the BNS simulations.
    The atmosphere and vacuum simulations use the same grid configurations
    to allow a proper comparison.
    $L$ denotes the total number of levels,
    $l^{mv}$ the finest non-moving level,
    $n$ $(n^{mv})$ the number of points in the fixed (moving) boxes, and
    $h_{0},h_{L-1}$ are the grid spacings in level $l=0,L-1$.
    The grid spacing of level $l$ is $h_l=h_0/2^l$.
    $n_r$ is the radial point number and $n_{\theta}$ is angular
    point number.}
  \begin{tabular}{|c|cccccccc|}
    \hline
     Resolutions  & $L$ & $l^{mv}$ & $n$ & $n^{mv}$ & $h_0$ & $h_{L-1}$ & $n_r$ & $n_{\theta}$\\
    \hline
     $\rm Low$ & 7 & 2 & 128 & 64 & 15.040 & 0.235  & 128 & 56\\
     $\rm Med$ & 7 & 2  & 192 & 96 & 10.027 & 0.157 & 192 & 84\\
     $\rm High$ & 7 & 2 & 256 & 128 & 7.520 & 0.117 & 256 & 112\\
     $\rm Fine$ & 7 & 2 & 320 & 160 & 6.016 & 0.094 & 320 & 140\\
     \hline
  \end{tabular}
 \label{Tab:tests-grid2}
\end{table}

\subsection{Binary configurations}

Finally, we want to discuss the performance of our new vacuum treatment
for the simulation of BNS setups.
We focus here on the simulation of an equal-mass, non-spinning configuration
evolved with the old atmosphere and the new vacuum method.
To save computational costs, we only perform simulations with the HOLLF scheme,
for which Ref.~\cite{Bernuzzi:2016pie} showed its superiority compared to
the LLF scheme with primitive reconstruction.

The individual stars have a baryonic mass of $1.495$ and
a gravitational mass in isolation of $1.350$.
For the EOS, we use a piecewise polytropic fit of the zero-temperature SLy
EOS~\cite{Chabanat:1997qh, Douchin:2001sv,Read:2008iy} and add an
additional thermal ideal-gas pressure component during the dynamical simulation.

The initial data is calculated by using the pseudo-spectral
SGRID code~\cite{Tichy:2006qn,Tichy:2009yr,Dietrich:2015pxa}.
The initial separation of the stars is $35.5$, i.e., $52.4\rm km$, which results
in an orbital frequency of $0.0070$, an initial ADM mass of $2.678$,
and an initial ADM angular momentum of $7.686$.
The eccentricity of the inspiral is approximately $1.3 \cdot 10^{-4}$.
This relatively low value has been achieved
by using the eccentricity reduction discussed in~\cite{Dietrich:2015pxa}.

Details about the grid setup for the BAM evolutions are given in
Tab.~\ref{Tab:tests-grid2}. To save computational costs, we have employed
bitant symmetry.
In contrast to the single-star tests, we substitute the outermost Cartesian
box (level $l=0$) by a shell made up from
six ``cubed sphere''
patches~\cite{Pollney:2009yz,Thornburg2000:multiple-patch-evolution,
Thornburg2004:multipatch-BH-excision}. In this shell matter is not evolved.

\subsection{Dynamical Evolution}
\label{sub-sec:dyn-evo}

\begin{figure*}[t]
\includegraphics[width=1.\textwidth]{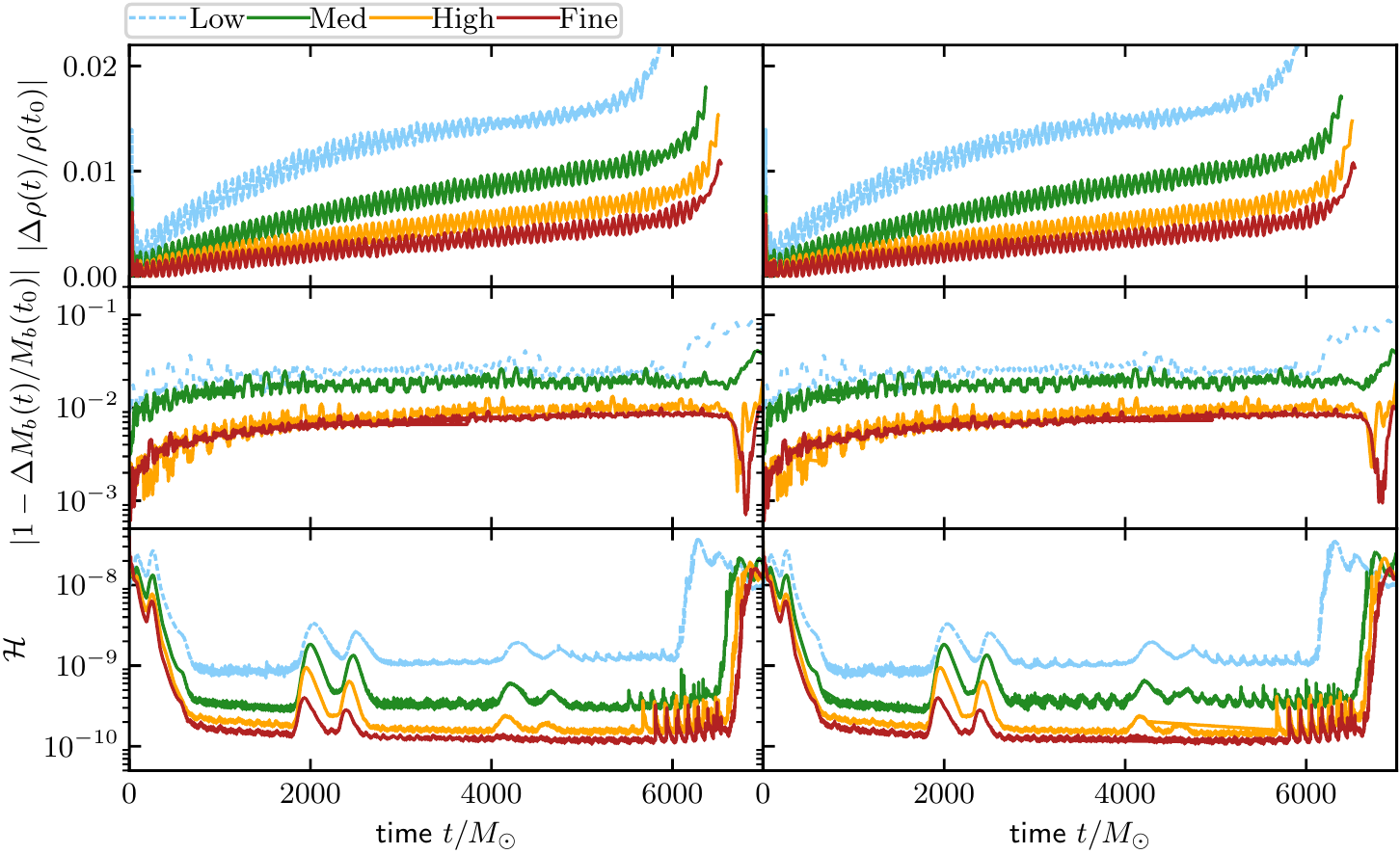}
\caption{Results of the BNS runs.
  Left to right: Atmosphere-HOLLF and Vacuum-HOLLF.
  Top: Relative change in central density $1 - \frac{\rho_c(t)}{\rho_c(t=0)}$.
  Middle: Relative rest-mass change $|1 - \frac{M_b(t)}{M_b(t=0)}|$.
  Bottom: The time evolution of Hamiltonian Constraint($\mathcal{H}$).
  We employ a Savitzky-Golay filter~\cite{Savitzky:1964a, Savitzky:1989a}
  to increase the visibility of the presented curves.}
\label{fig:test-bns}
\end{figure*}

\begin{figure}[t]
\includegraphics[width=0.5\textwidth]{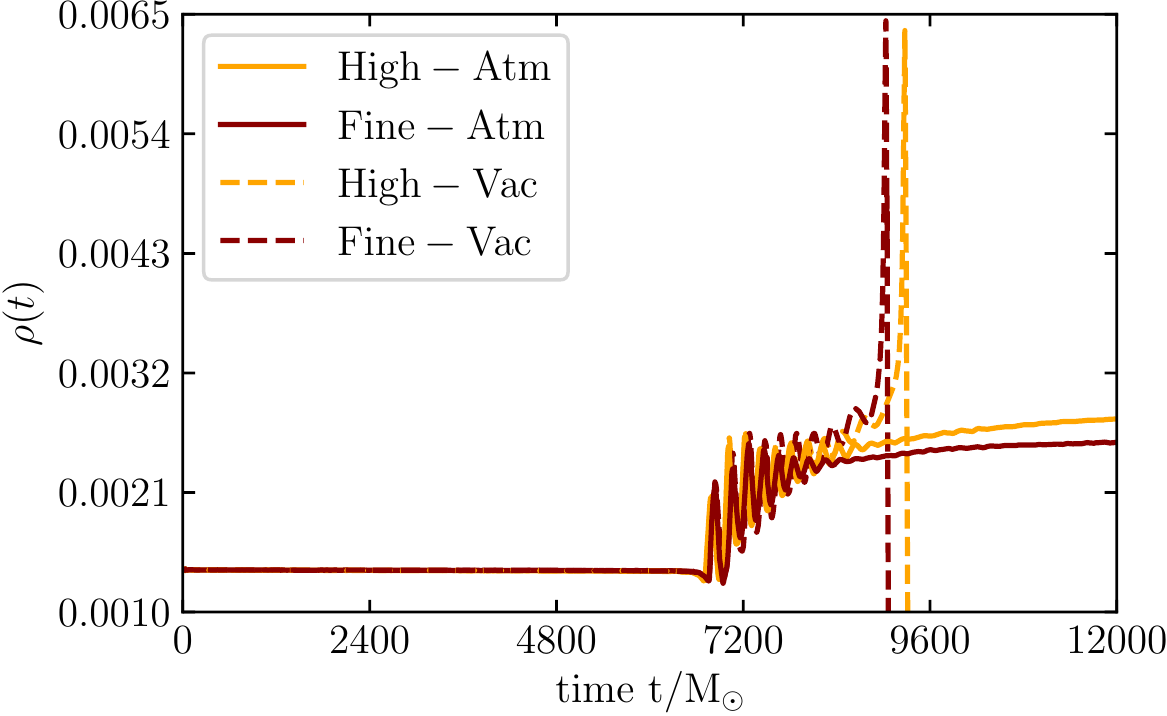}
\caption{Central density of the BNS for the simulations with
the two highest resolutions. Solid lines
are for the atmosphere and the dashed lines are for the vacuum methods. This
plot shows how density changes in BNS during inspiral, merger, post-merger,
and black hole formation.}
\label{fig:bns-density}
\end{figure}

During the inspiral, a general (almost linear)
reduction of the central density is visible.
This linear trend reduces significantly with increasing resolution and is
connected to the numerical dissipation~\cite{Bernuzzi:2016pie},
which decreases with decreasing grid spacing; cf.\ Fig.~\ref{fig:test-bns}.
Overall, there is generally a second-order convergence in the central
density visible for both, the atmosphere and the vacuum method.

In addition, we plot the time evolution for the central density for the
two highest resolutions in Fig.~\ref{fig:bns-density}.
Clearly visible are large density oscillations after the merger,
which correspond to radial oscillations of the formed hypermassive neutron star (HMNS),
see e.g.\ Refs.~\cite{Stergioulas:2011gd,Rezzolla:2016nxn} for further details.
The main difference between the old atmosphere and the new vacuum method is
that for the two highest resolutions the lifetime of the HMNS is shorter for the vacuum method
than for the atmosphere treatment. We note that the determination of the remnants lifetime
does influence (i) the material outflow and its composition and (ii) the properties of the BH+disk
system, i.e., the potential short gamma-ray burst.

The middle panel of Fig.~\ref{fig:test-bns} shows the conservation of the
rest-mass density, where we note that these simulations do not employ
the conservative refluxing algorithm yet. We plan to repeat the simulations
with conservative refluxing in the future when we have more computer time.
The error of the rest-mass seems to decrease as we go to higher
resolutions. For the highest resolution, the total rest-mass in conserved up to
$0.5\%$ throughout the inspiral, independent of the employed atmosphere/vacuum scheme.
For the Hamiltonian constraint, convergence consistent with second-order
is seen until $2000M_{\odot}$ in both the atmosphere and vacuum case. After
that, the order of the convergence rises up to fourth order
which is is higher than is theoretically expected.
However, throughout the simulation there is clear pattern of Hamiltonian
constraint decreasing for both cases as we go to higher resolutions.
Looking at the plots of these three quantities, there is not a clear advantage
for either the atmosphere or the vacuum method.

\begin{table}[t]
  \centering
  \caption{Merger times for atmosphere and vacuum cases at different
   resolutions. Here Low, Med, High, and Fine are simulation with
   resolutions 64, 96, 128, and 160 points respectively. The merger time values
   are in geometric units where $6000M_{\odot} \approx 30ms$.}
  \begin{tabular}{|l|cccc|}
    \hline
     Tests & $Low$ & $Med$ & $High$ & $Fine$  \\
    \hline
     BNS Atm & $6117M_{\odot}$ & $6583M_{\odot}$  & $6682M_{\odot}$ & $6740M_{\odot}$  \\
    \hline
     BNS Vac & $6150M_{\odot}$ & $6612M_{\odot}$  & $6689M_{\odot}$ & $6737M_{\odot}$  \\
     \hline
  \end{tabular}
 \label{Tab:BNS:mergertime}
\end{table}

\subsection{Gravitational Waveforms}
\label{sub-sec:GW}

\begin{figure*}[t]
\includegraphics[width=0.95\textwidth]{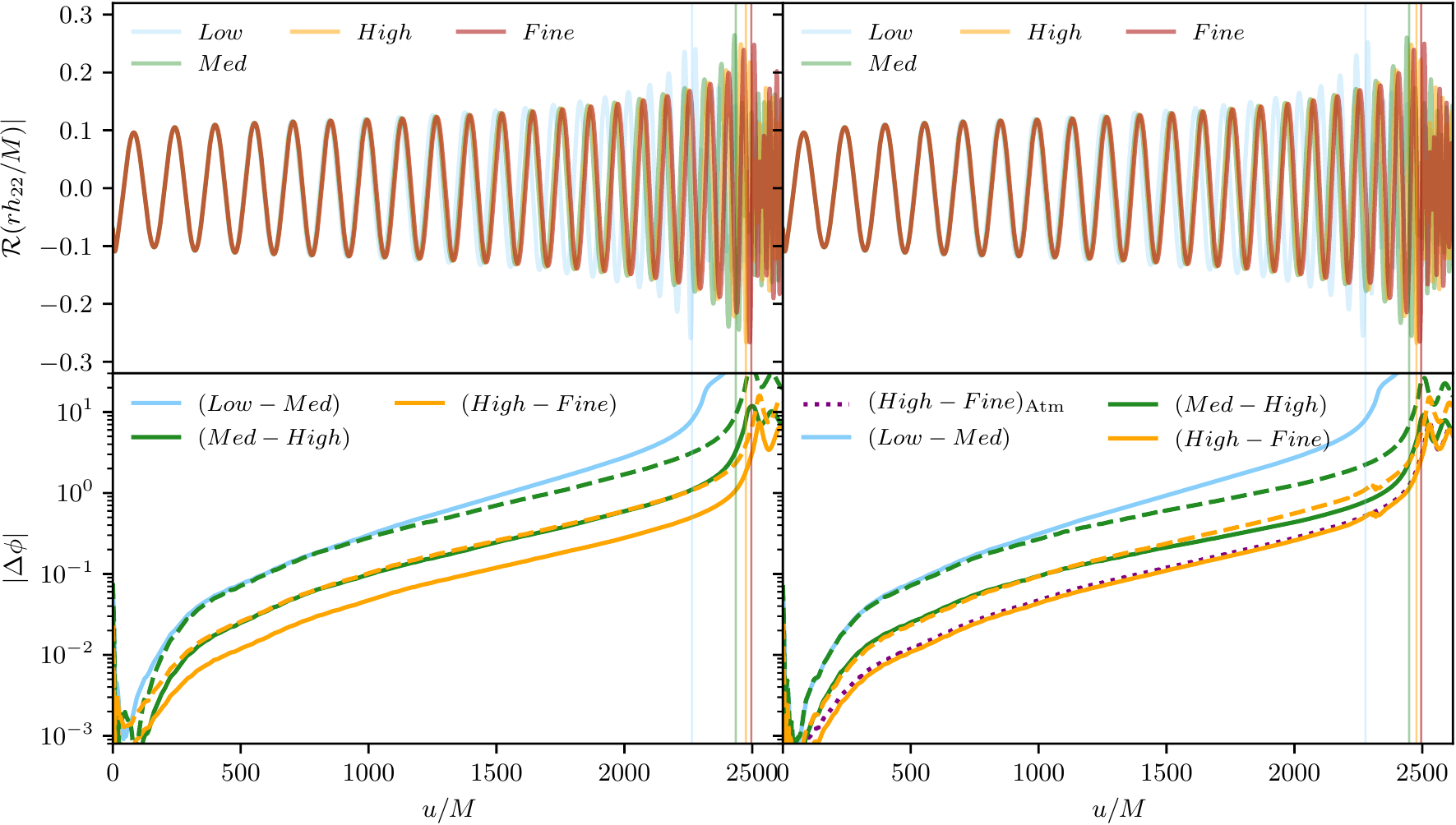}
\caption{In the top panels, we plotted the GW signal for both the old
atmosphere method (left) and the new vacuum method (right). The bottom panels show
the phase differences between different resolutions. The dashed lines are
the rescaled phase differences assuming second-order convergence. The bottom
right panel includes also the phase difference of the two highest resolutions 
for the atmosphere method for an easier comparison. 
Overall, we find similar phase difference and convergence properties for methods.
}
\label{fig:GW}
\end{figure*}

For both methods an increasing resolution leads to a later merger, which is
due to the increase of the numerical dissipation for lower resolutions. For
simulations with the BAM code this effect is discussed
in~\cite{Bernuzzi:2016pie}.
We report the merger time in Tab.~\ref{Tab:BNS:mergertime} for all BNS simulations.
Most importantly, the difference between the atmosphere and vacuum method decreases
with increasing resolution, consequently, both methods seem to lead to a
similar continuum limit.

In Fig.~\ref{fig:GW}, we present the GW signal for all
simulations in the top panels for the atmosphere (left) and the vacuum
(right) methods. The bottom panels shows the phase differences
between different resolutions, we rescale the phase differences assuming second-order
convergence (dashed lines) and find generally that the both methods show the expected
convergence order with a sightly better convergence behavior for the
original atmosphere treatment. For all methods the $\rm Low$-resolution
simulation stops being second-order convergent after about $1500 M$, which
indicates that this resolution is not sufficient to be in the convergent regime until merger.
Overall, we find very close agreement between the individual phase differences reported in
the Fig.~\ref{fig:GW}. In the right bottom panel, we show the phase difference between the two highest
resolutions for the vacuum (solid, orange) and the atmosphere method (purple, dotted).
We find that the difference is almost identical, thus, 
there is no improvement in the extracted GW signal for our new vacuum method,
which we assume is caused by the fact that the overall bulk motion
is dominating the GW radiation and that the new vacuum treatment mostly
affects the low density regions.

\subsection{Ejecta Quantities}
\label{sub-sec:ejecta}

\begin{figure}[t]
\includegraphics[width=0.5\textwidth]{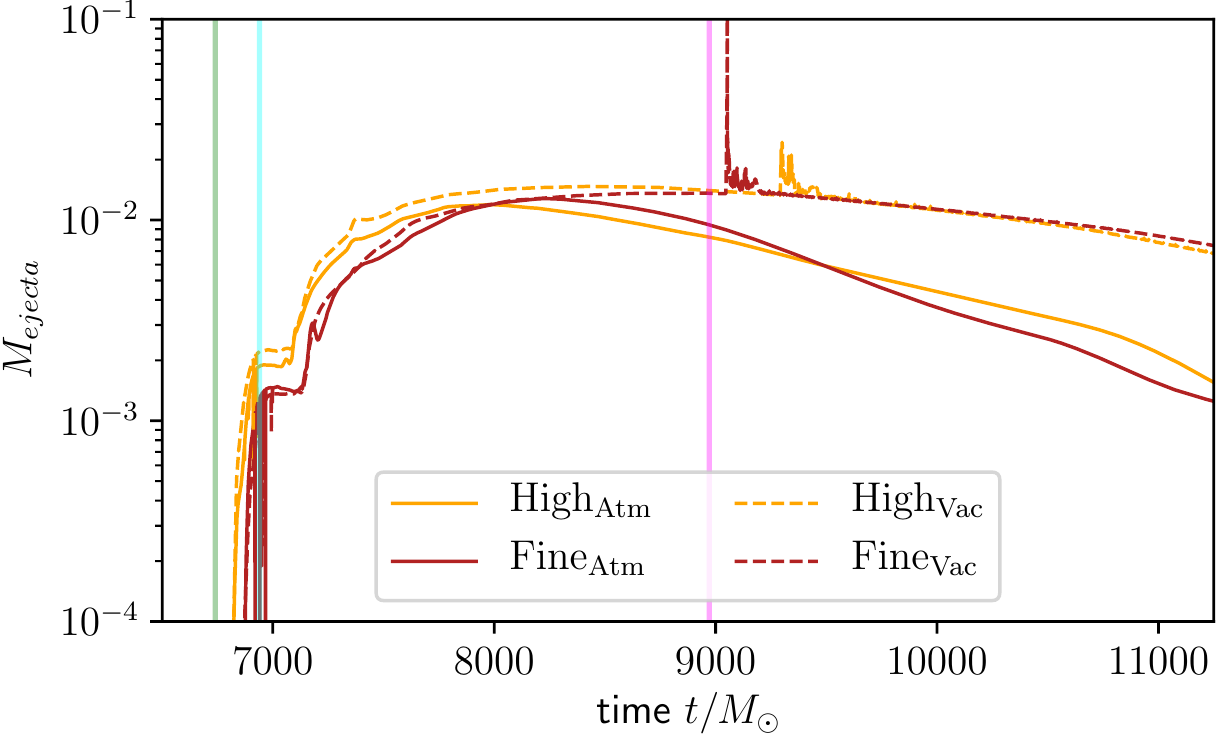}
\caption{The ejecta mass for the two highest resolutions. The vertical lines indicate the merger time,
the time $1ms$ after the merger, and the time $11ms$ after the merger.
For the latter two, 2D plots are shown in Figs.~\ref{fig:ejecta-2D1} and
\ref{fig:ejecta-2D2}. The spike in the vacuum simulations occurs due to
the formation of the BH. }
\label{fig:ejecta}
\end{figure}

\begin{figure*}[t]
\includegraphics[width=0.45\textwidth]{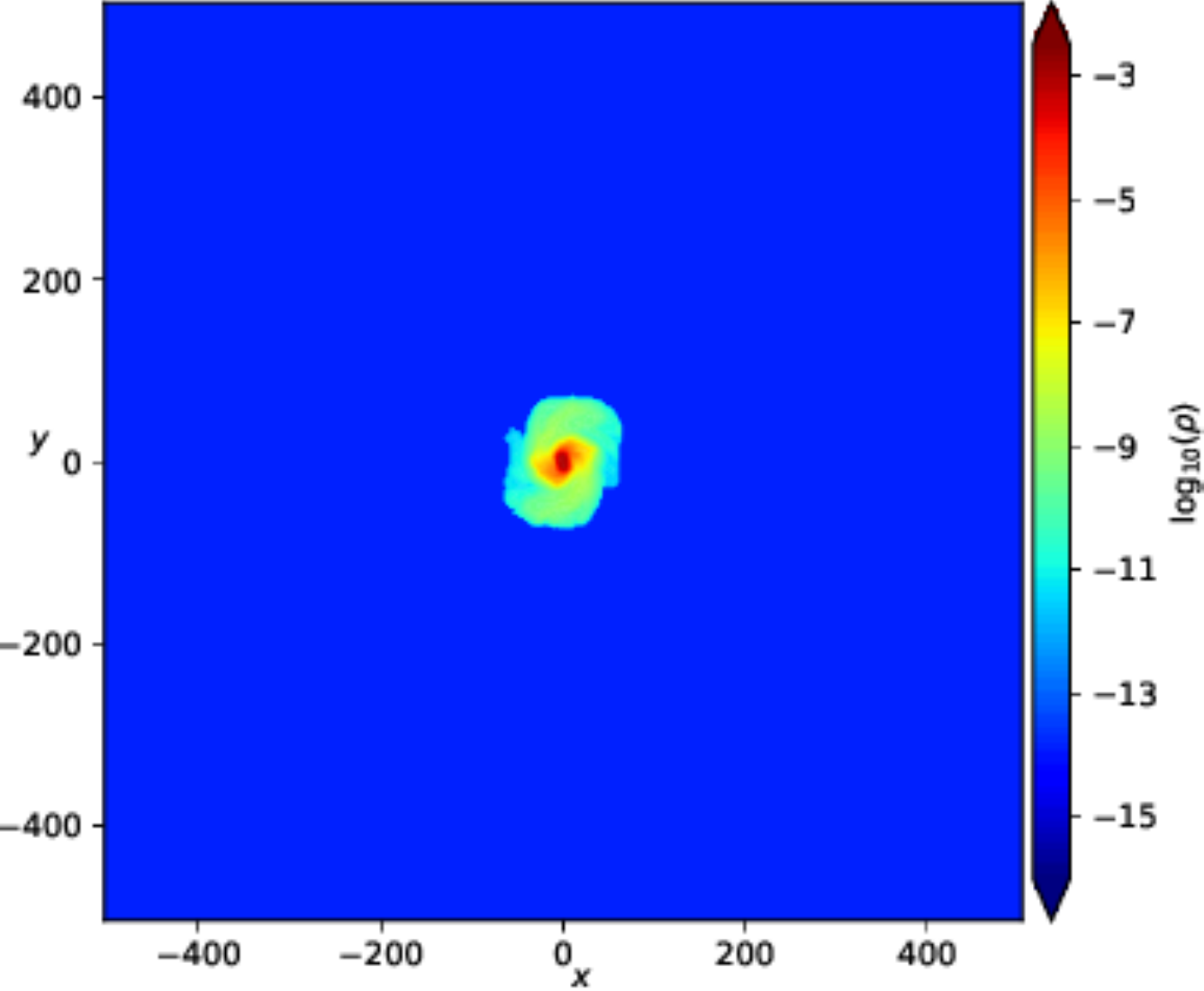}\qquad
\includegraphics[width=0.45\textwidth]{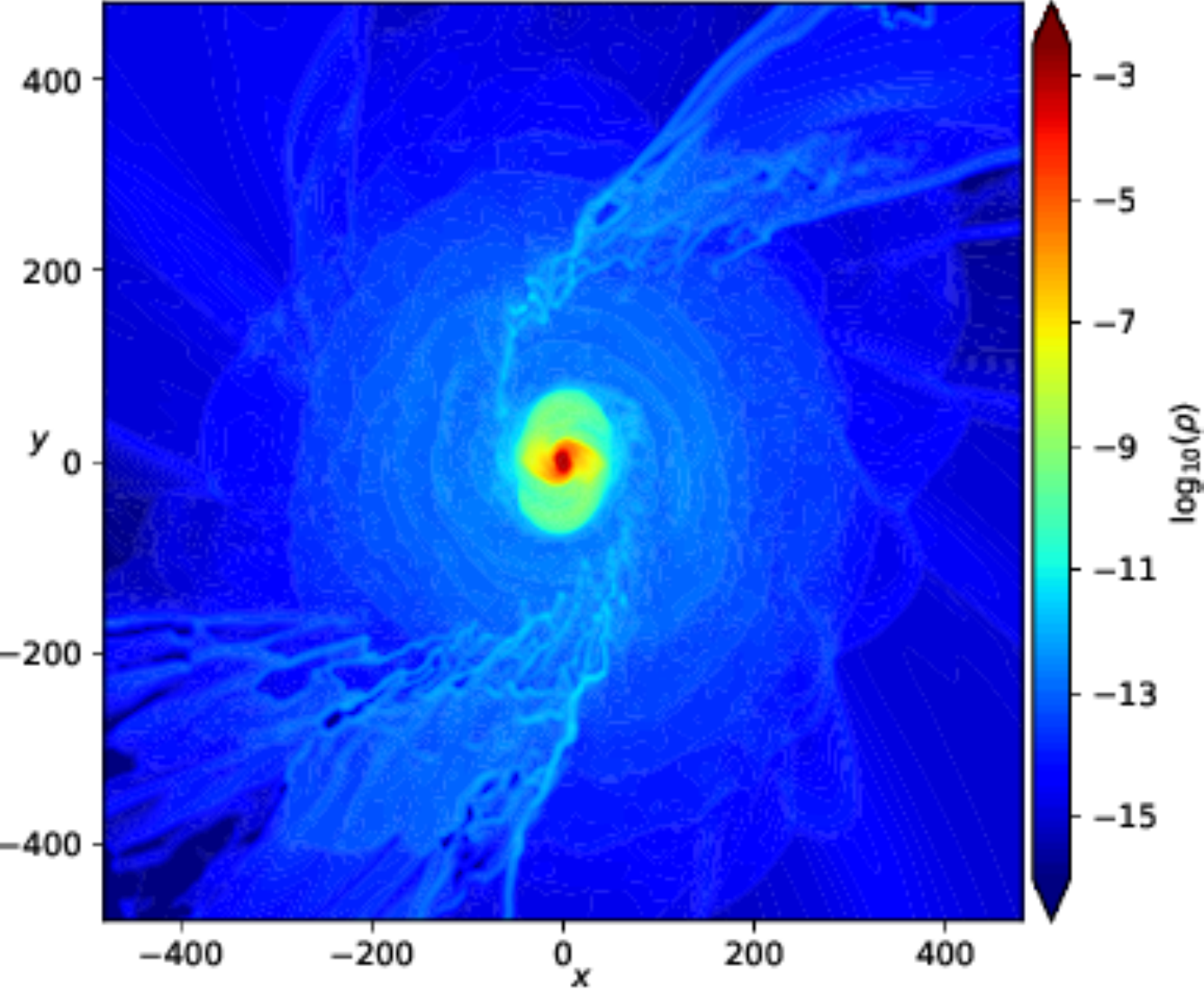}\\
\includegraphics[width=0.45\textwidth]{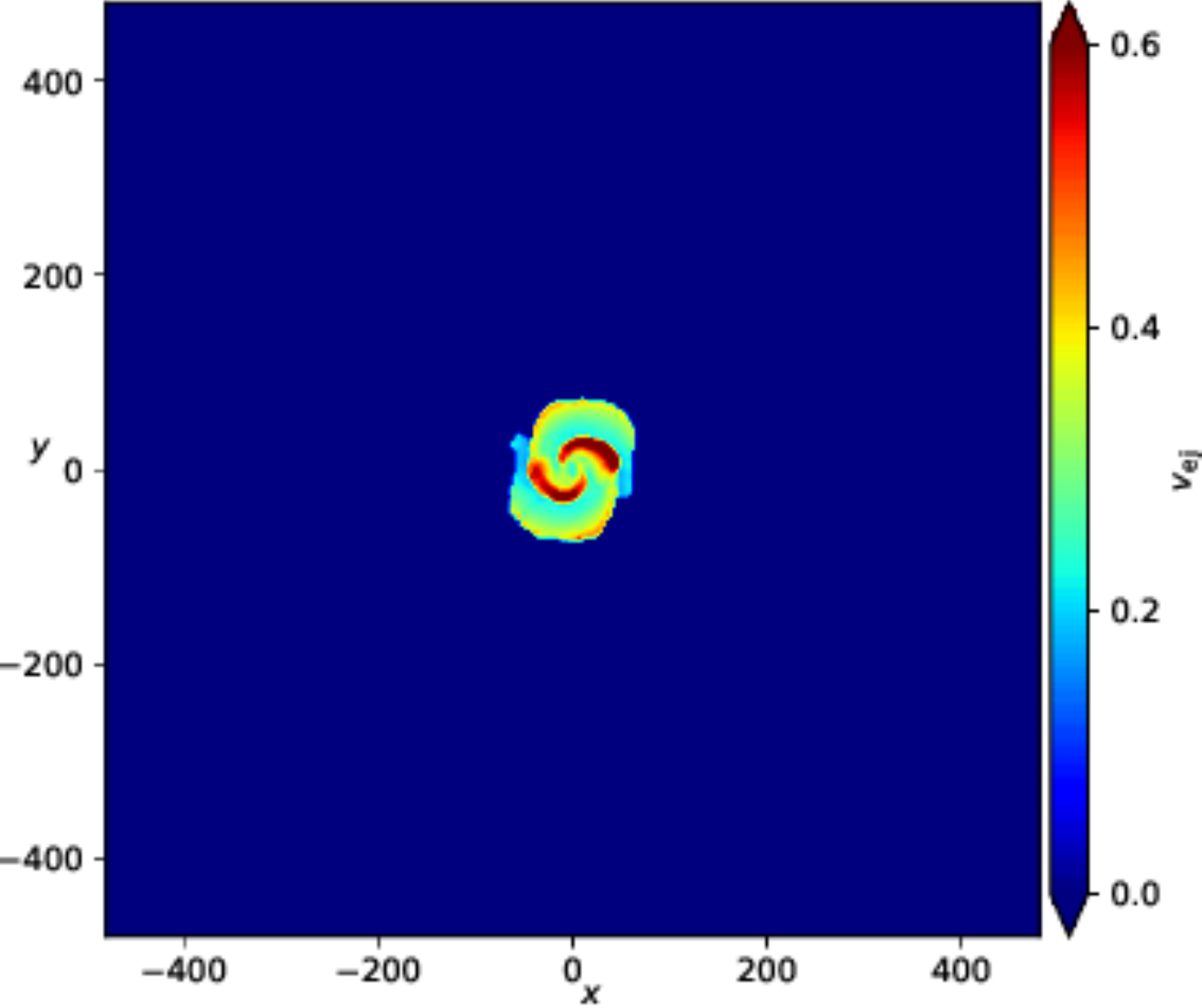}\qquad
\includegraphics[width=0.45\textwidth]{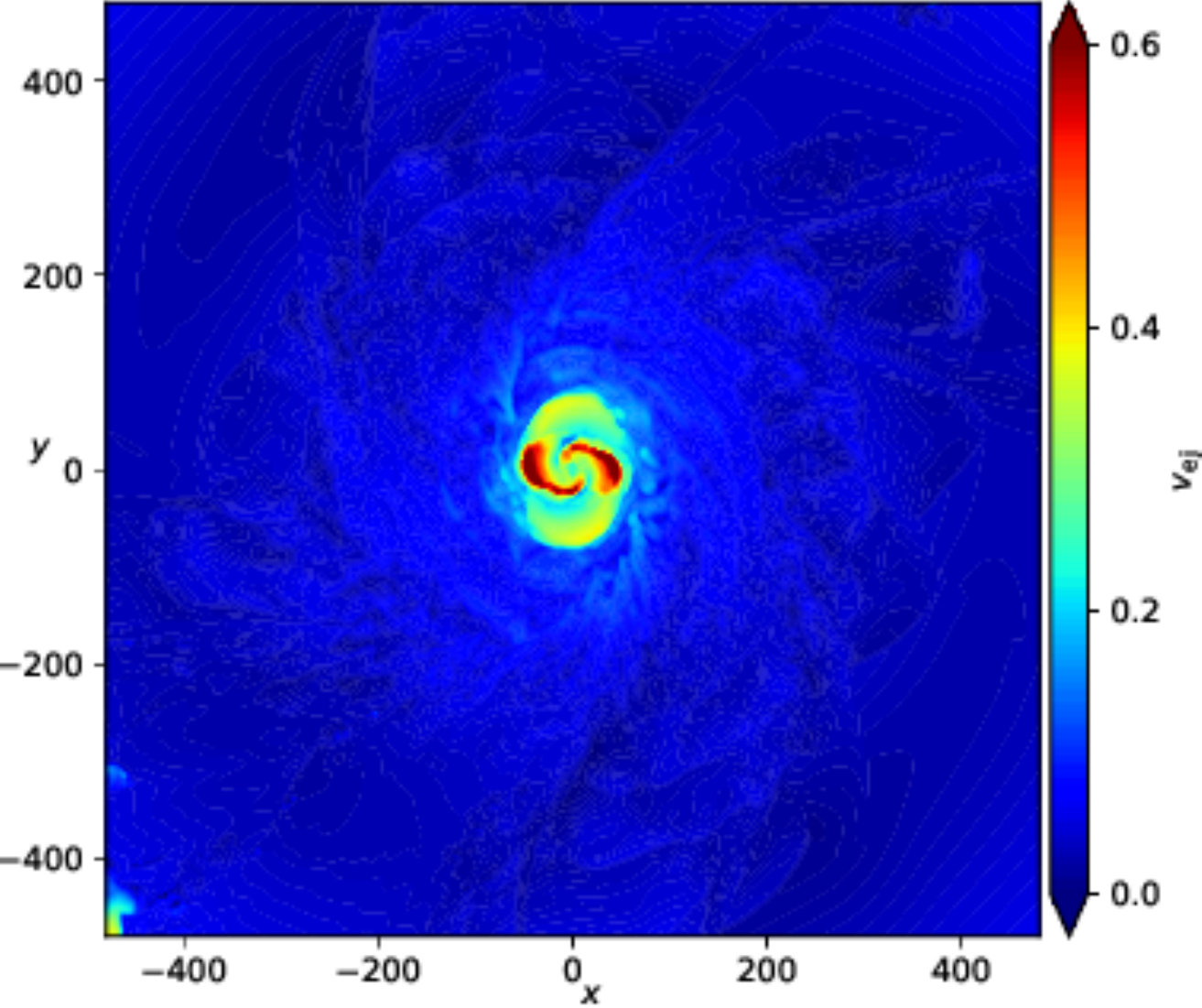}\\
\includegraphics[width=0.45\textwidth]{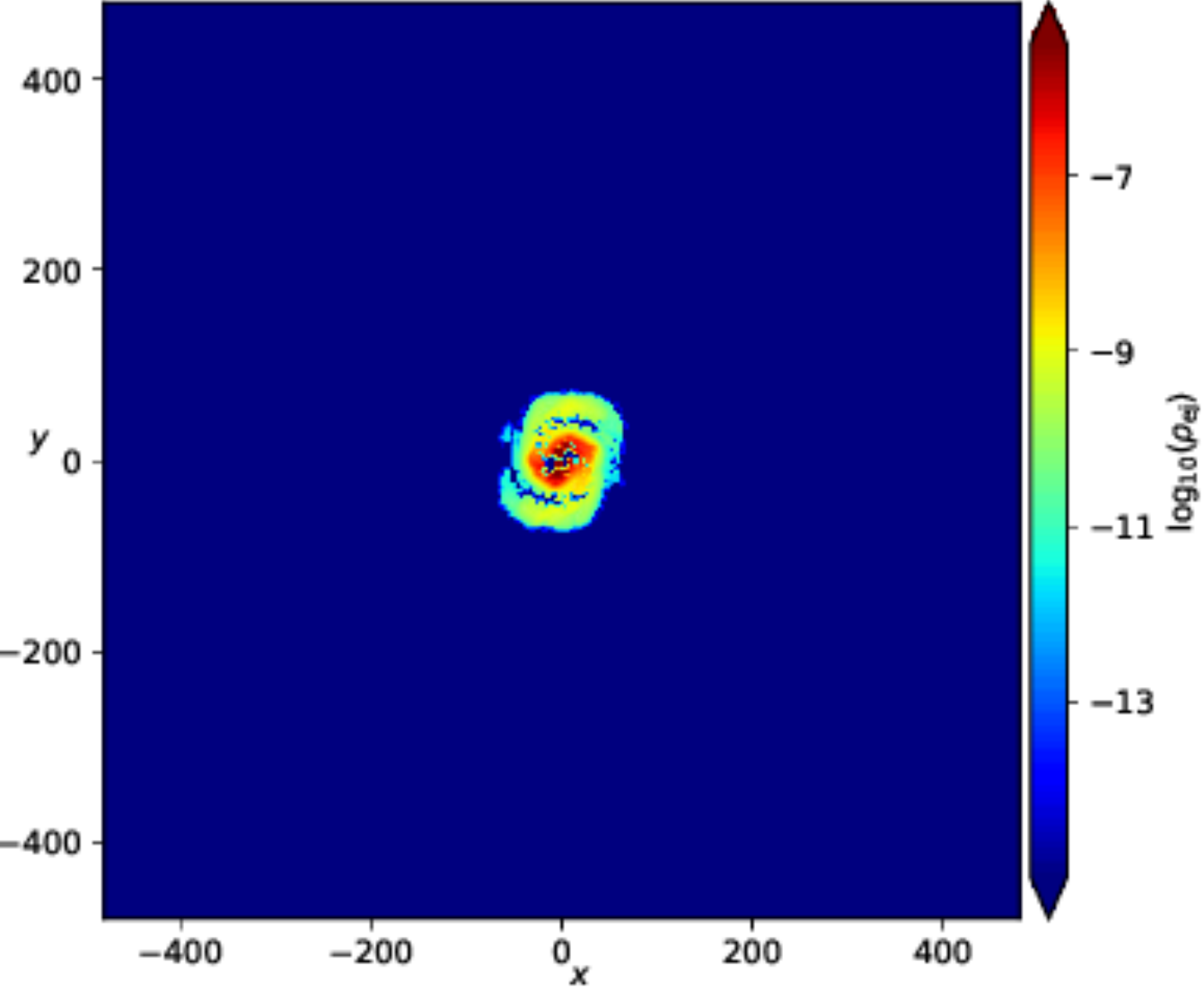}\qquad
\includegraphics[width=0.45\textwidth]{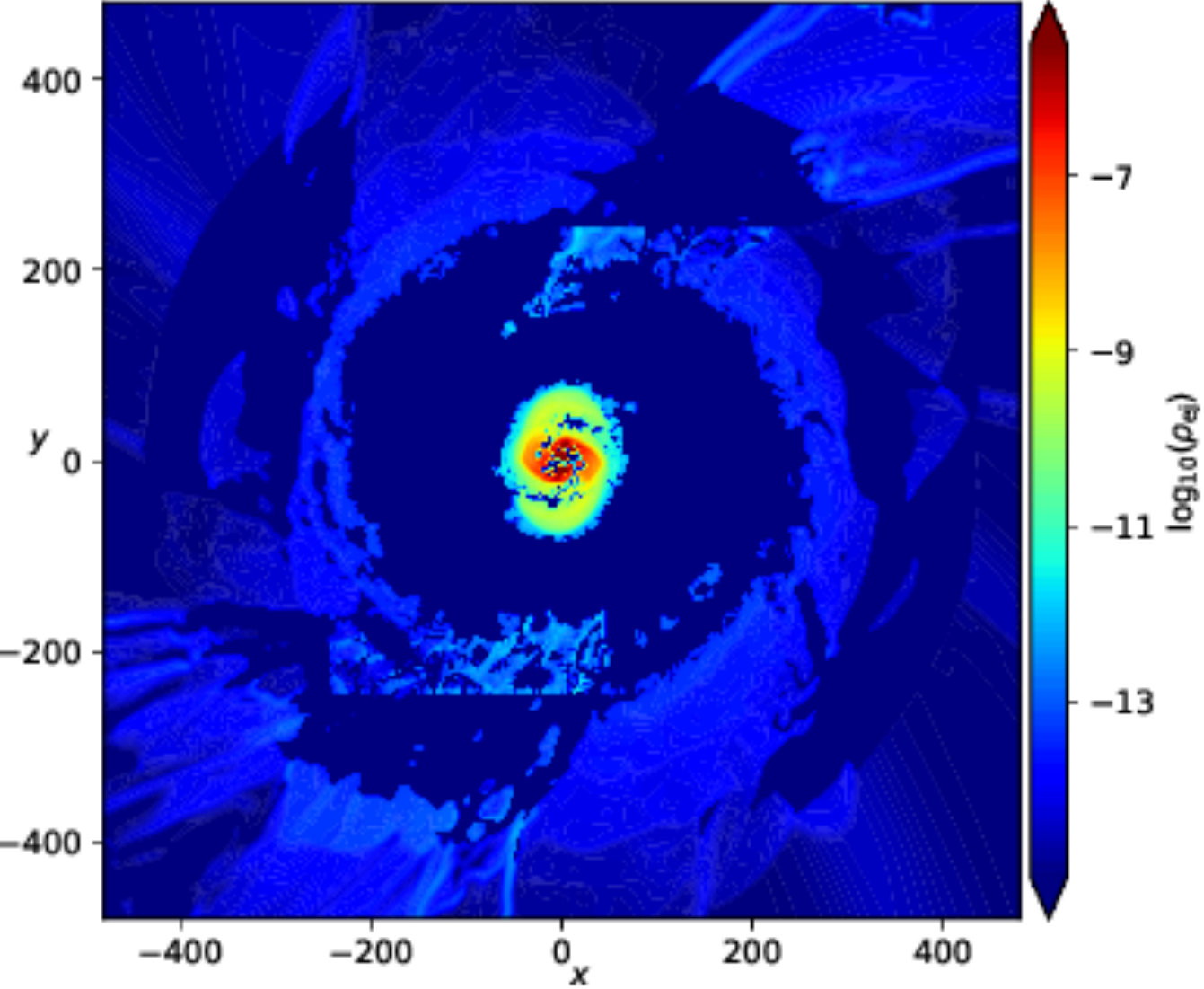}\\
\caption{Snapshots of the mass density (top), velocity (middle), and ejecta mass
density (bottom) in xy-plane of BNS simulation at $1ms$ after
merger (vertical cyan line in Fig.~\ref{fig:ejecta}).
The finest resolution for
both the atmosphere (left) and the vacuum (right) is
plotted with a linear (for velocity) and logarithmic color scales
(for mass density and ejecta mass density). This is on level $l=1$,
which extends up to 481.28. The atmosphere threshold density
is $\rho_{thr} = 1.389 \cdot10^{-12}$. We label material as unbound in
case the fluids 0th component of the 4-velocity is smaller than one $u_t<-1$
and if the three velocity is radially outward pointing.}
\label{fig:ejecta-2D1}
\end{figure*}

\begin{figure*}[t]
\includegraphics[width=0.45\textwidth]{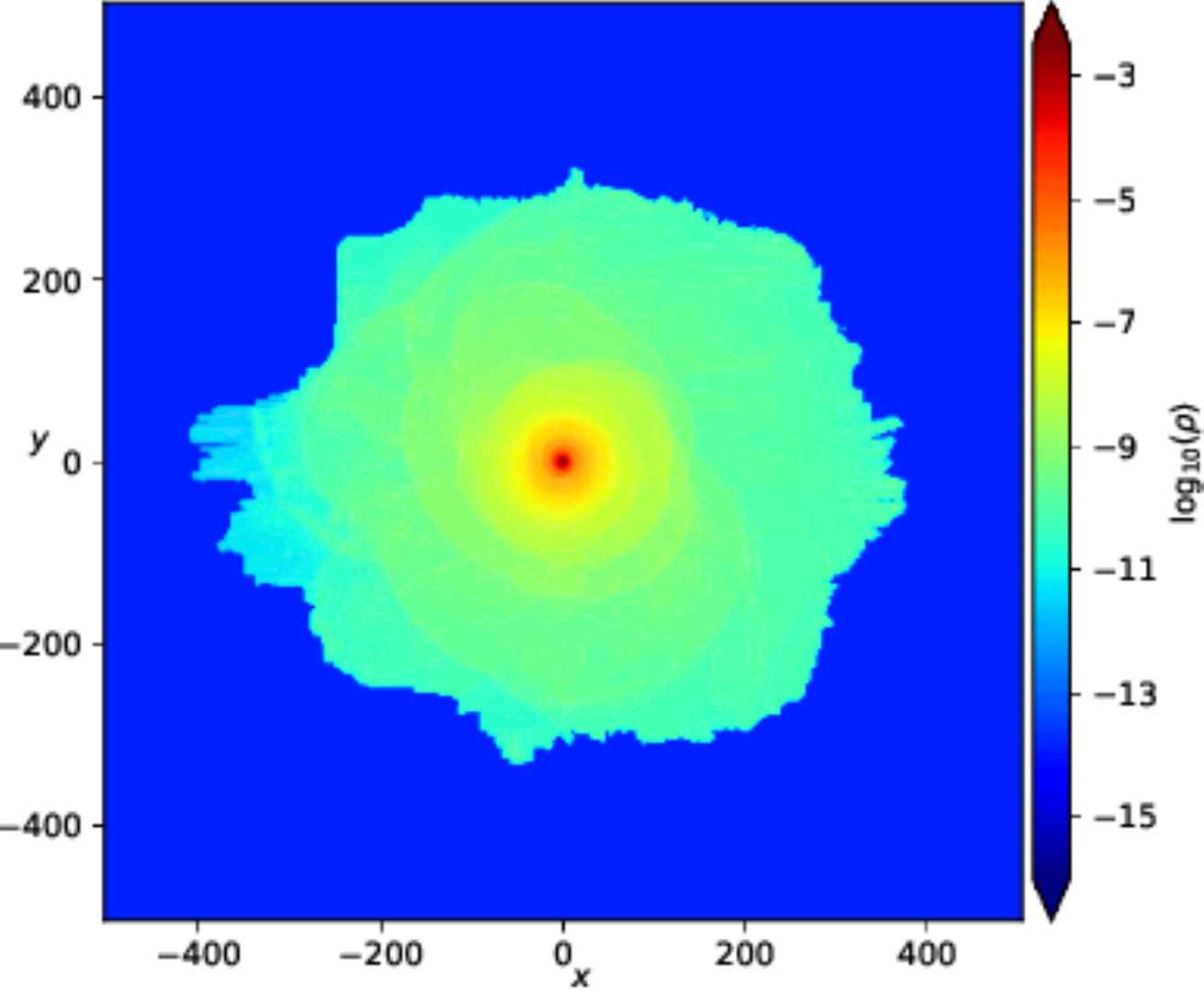}\qquad
\includegraphics[width=0.45\textwidth]{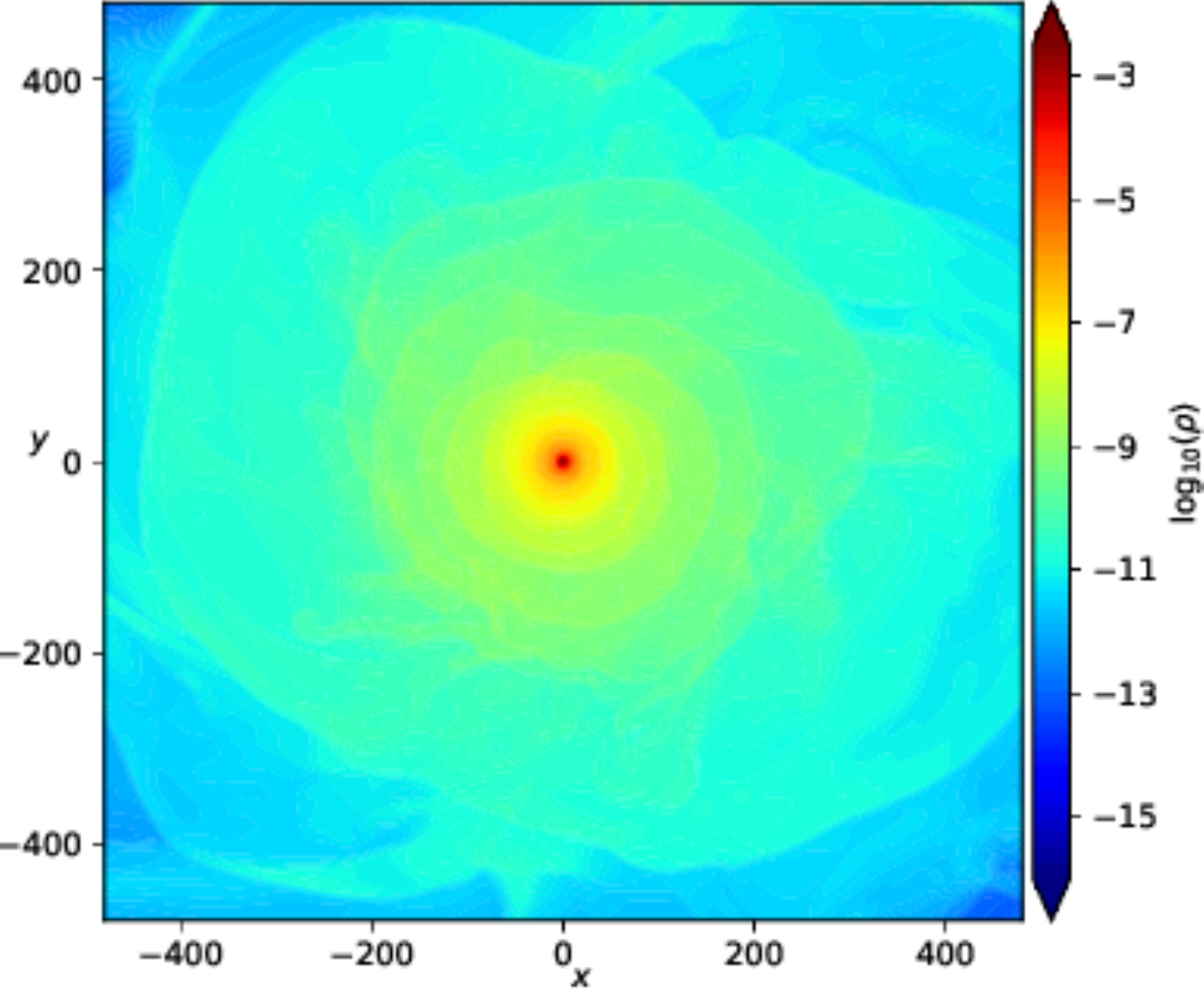}\\
\includegraphics[width=0.45\textwidth]{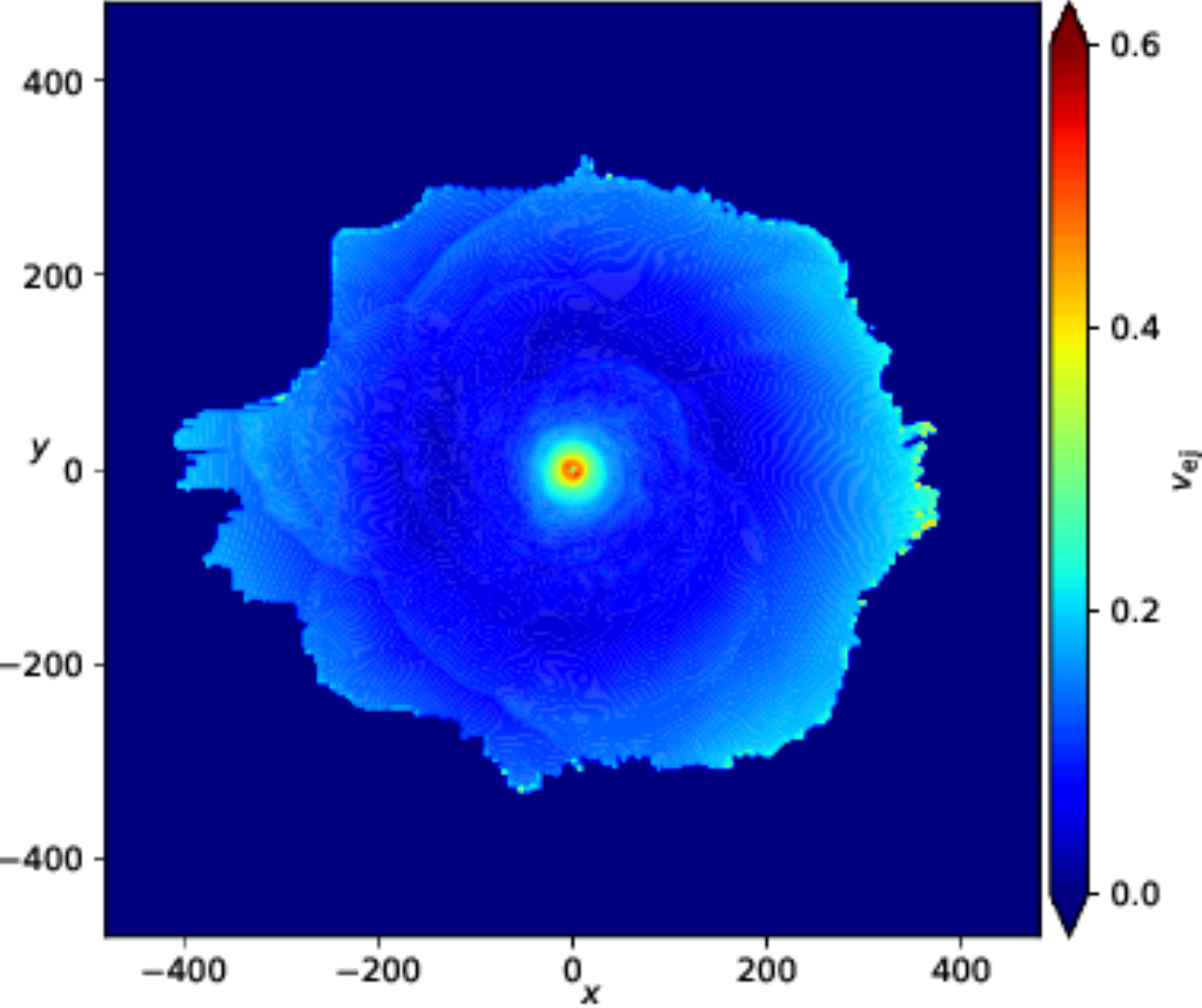}\qquad
\includegraphics[width=0.45\textwidth]{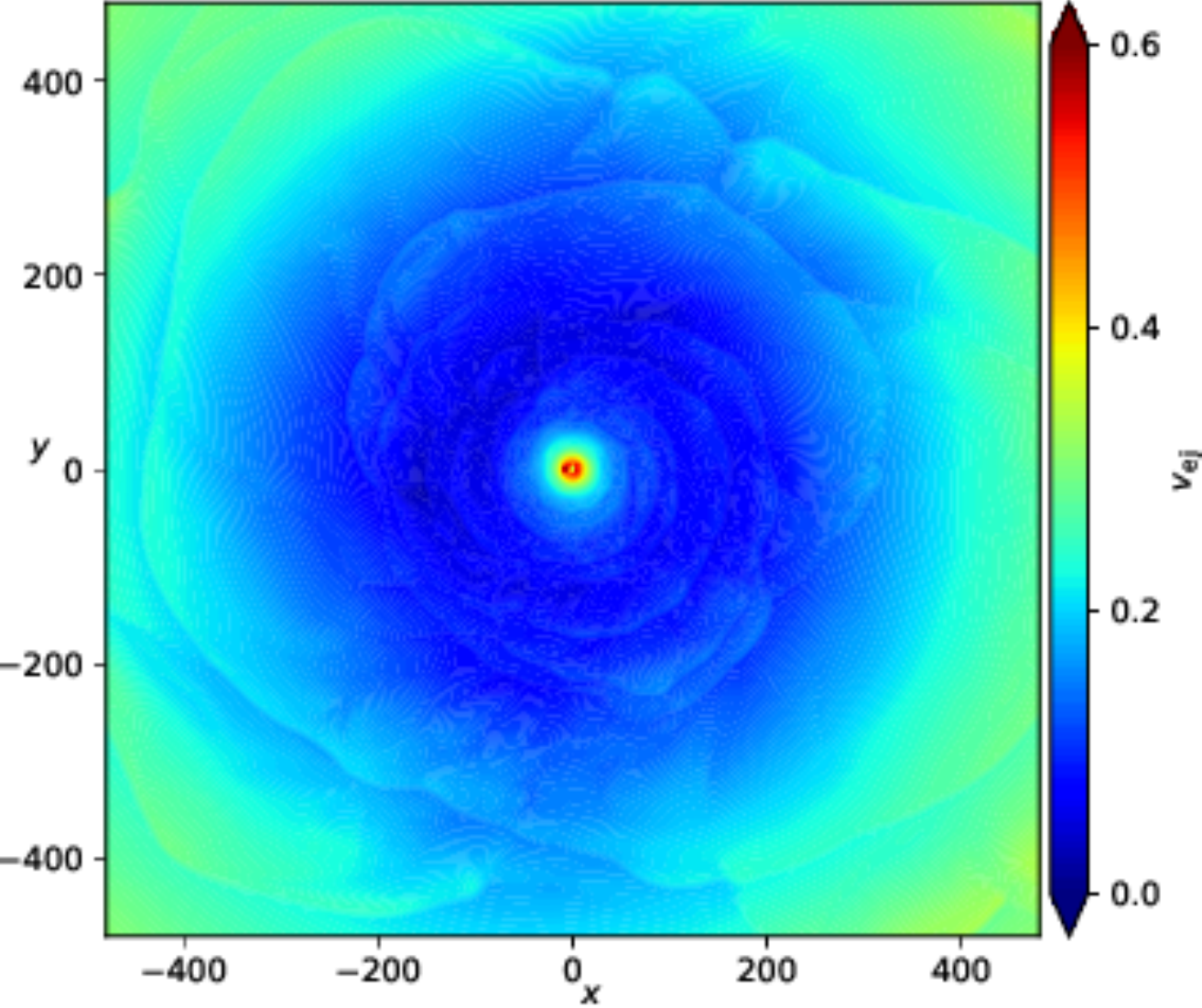}\\
\includegraphics[width=0.45\textwidth]{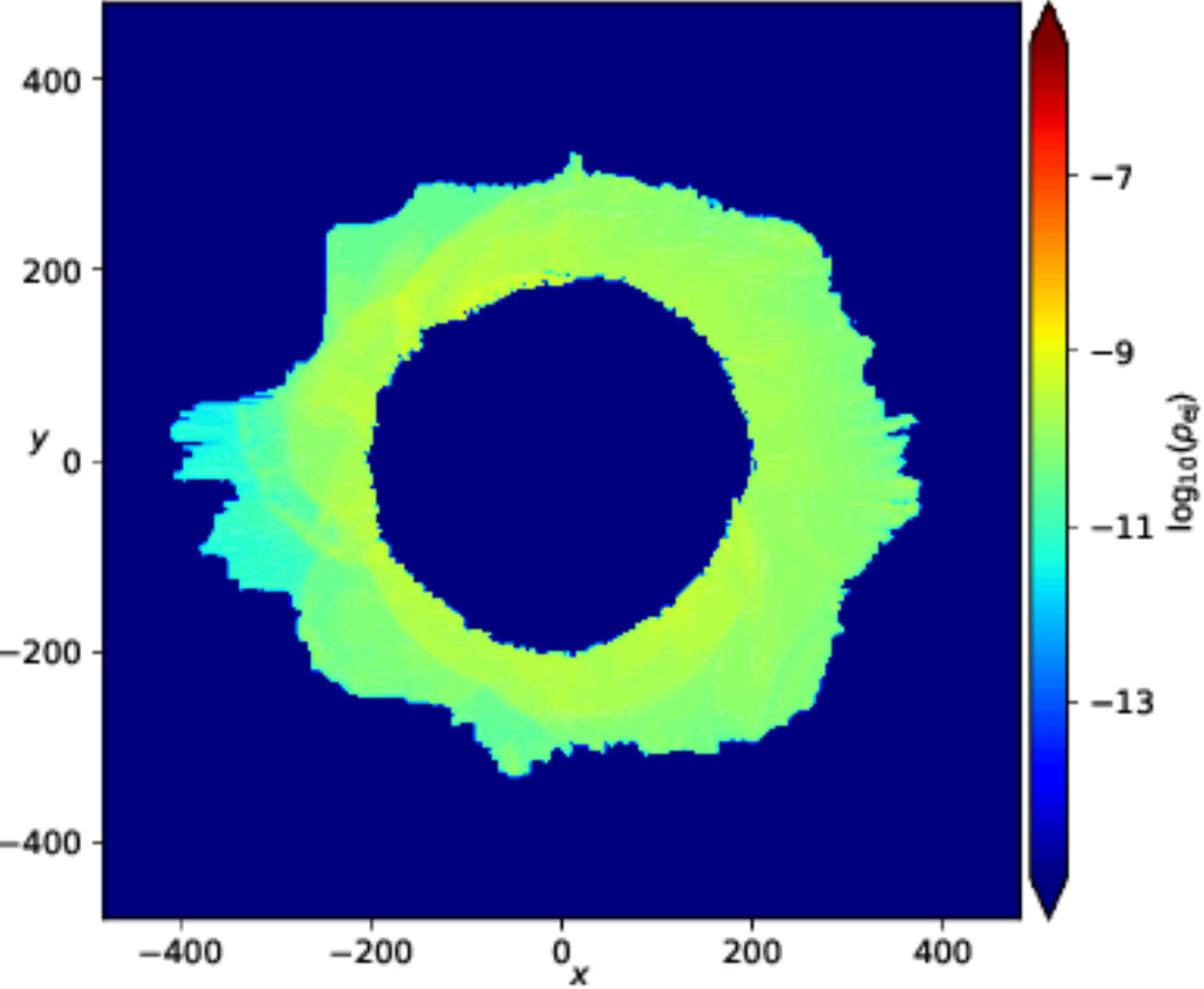}\qquad
\includegraphics[width=0.45\textwidth]{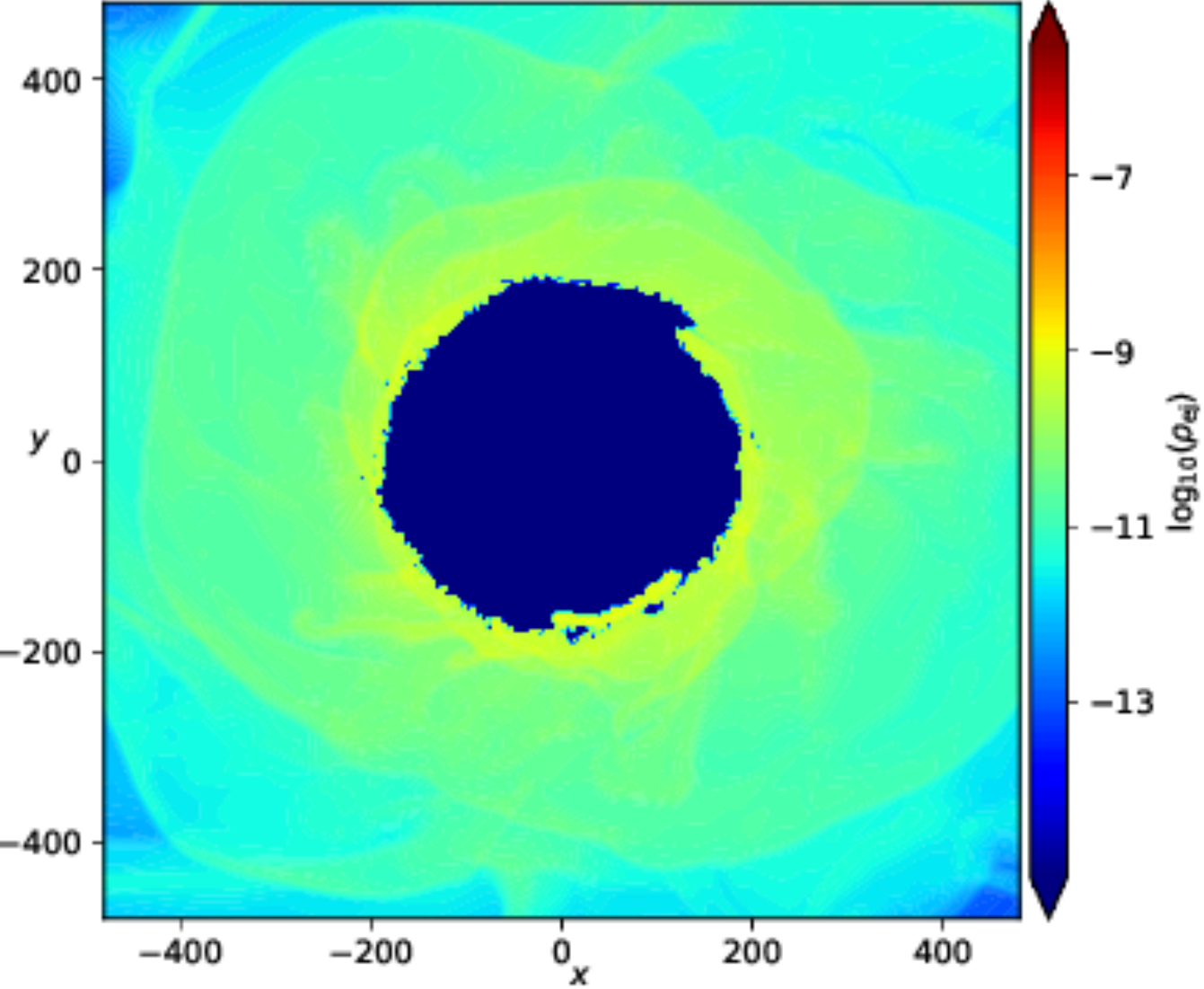}\\
\caption{Snapshots of the mass density (top), velocity (middle), and ejecta mass
density (bottom) in xy-plane of BNS simulation. The finest resolution for
both the atmosphere (left) and the vacuum (right) at $11ms$ after the
merger (vertical cyan line in Fig.~\ref{fig:ejecta}) is
plotted with a linear (velocity) and logarithmic (mass density and ejecta mass
density) color scales.}
\label{fig:ejecta-2D2}
\end{figure*}

Since the amount of ejected material is tightly connected to
the creation of electromagnetic counterparts for a BNS merger,
see e.g.\ Ref.~\cite{Metzger:2016pju} and references therein,
it is important to improve the evolution of low density material.

Generally, when material gets ejected, the fluid expands and the density
reduces until it finally falls below the artificial atmosphere threshold
used within the atmosphere implementation. In the atmosphere case
such fluid elements are then set to atmosphere values with
zero velocity, and are thus no longer
counted as ejecta, so that the ejecta mass decreases. This trend
is clearly visible in Fig.~\ref{fig:ejecta}.
The problem is mostly present at the outer region of the ejected material
for material which moves with the highest velocities.
Consequently, removing this material leads also to a drop in
$v_{\rm ej}$ for the atmosphere simulations.

We have made 2D plots to further investigate the difference in
mass ejection in the atmosphere and the vacuum methods
(see Figs.~\ref{fig:ejecta-2D1} and~\ref{fig:ejecta-2D2}).
We plot the mass density, velocity, and ejecta mass density
at level $l=1$ of the finest resolution simulation.
We use a linear (velocity) as well as a logarithmic (mass density and
ejecta mass density) color scales and plot two snapshots in time at $1ms$ (Fig.~\ref{fig:ejecta-2D1})
and $11ms$ (Fig.~\ref{fig:ejecta-2D2}) after the merger. We choose these two times for the
following reasons. Both methods begin with identical ejecta, same
velocity and similar mass density. As the ejecta expand to a larger
radius the
density in the outer regions of the ejecta mass drop below the atmosphere
threshold value. Thus, low-density material is removed and the ejected
matter never reaches the outer boundary.
Therefore, we find that the outflowing material seems to stall
about $\sim 10ms$ after the merger at a maximum radial extend of $\sim 350$.
In the vacuum case, the ejecta moves further out.
The eventual mass reduction for the vacuum method in Fig.~\ref{fig:ejecta}
at late times, can be explained by unbound material
reaching the boundary of level 1, which is the outermost refinement
level where matter is evolved.
We also note that we see in the bottom panel right panel of
Fig.~\ref{fig:ejecta-2D1} a clear imprint of the refinement boundaries on
the low density material.
We expect that, as in the TOV$_{\rm mig}$ case, the
use of the conservative refluxing algorithm would resolve this issue, but
postpone this test due to the high computational costs for the presented
BNS simulations.

\section{Conclusion}
\label{sec:conclusion}
In this article, we have introduced and studied a new method to improve
the vacuum treatment for GRHD simulations.
Our recipe allows to not set an explicit atmosphere value on the outside of a star which
improves the quality of our simulations.
Previously, we have implemented a method in the BAM code that used an artificial
atmosphere while recovering primitive variables. We extensively tested
both methods (vacuum and atmosphere) in single star spacetimes
focusing on their performance when combined with a second order local Lax-Friedrich and
higher-order numerical flux schemes.
The use of vacuum methods shows improvement in the mass conservation throughout our simulations.
Typically, for the star that forms a low-density region during evolution, the
mass conservation drops in the atmosphere method. Up to $0.5\%$ mass loss
was detected in the atmosphere method when the low-density
layer crosses the refinement boundary. The
violation of mass conservation at the grid refinement boundary does not
occur in the case of our improved vacuum method. In most cases, the
vacuum method leads to second-order convergence of the mass,
in contrast to the atmosphere case, where often second-order convergence is only
obtained for a short period of time.
Our finding suggests that the use of the vacuum method is desirable and
recommended for the single star simulations.

To further investigate the performance of the new vacuum method we presented
time evolutions of irrotational equal mass binary
neutron star configuration. Mainly, the merger and the postmerger dynamics
are of great interest because the artificial atmosphere setup hinders the
accurate computation of the ejecta~\cite{Dietrich:2015iva}.
Our analysis suggests that the ejecta materials are better conserved
with the vacuum method.
Around the moment of merger, the ejected mass,
the ejecta velocity, and the kinetic energy of the ejecta
are within the same range for both
methods (vacuum and atmosphere) for all resolutions.
But the difference in those quantities becomes prominent as
the ejecta expands to larger radii, and the density of the ejecta drops
below the atmosphere threshold. In the atmosphere cases this leads
to ejecta removal and does not allow a free expansion of the ejecta material.
In contrast, ejected matter can expand freely for the vacuum method.


\begin{acknowledgments}
  It is a pleasure to thank S.~Bernuzzi and S.~V.~Chaurasia for fruitful discussions
  during this project.
  W.T. was supported by the National Science Foundation under grants
  PHY-1305387 and PHY-1707227.
  B.B. was supported by DFG grant BR-2176/5-1.
  Computations were performed on SuperMUC at the LRZ (Munich) under
  the project number pr46pu and pn56zo, Jureca (J\"ulich)
  under the project number HPO21, and Stampede
  (Texas, XSEDE allocation - TG-PHY140019).
\end{acknowledgments}


\bibliography{references}

\end{document}